\documentclass[twocolumn]{aastex61}
\usepackage{amsmath,amstext}
\usepackage{amssymb}
\usepackage{apjfonts} 
\usepackage{graphicx}
\usepackage[figure,figure*]{hypcap}
\usepackage[version=3]{mhchem}
\usepackage{mysymbol}
\usepackage{savesym}
\savesymbol{tablenum}
\usepackage{siunitx}
\restoresymbol{SIX}{tablenum}
\usepackage{threeparttable}
\usepackage{tablefootnote}
\usepackage{scalefnt}
\usepackage{comment}
\usepackage{appendix}
\usepackage{url}
\usepackage[flushmargin]{footmisc}
\usepackage{bibspacing}
\def\simge{%
    \mathrel{\rlap{\raise 0.511ex
    \hbox{$>$}}{\lower 0.511ex \hbox{$\sim$}}}}
\def\simle{%
    \mathrel{\rlap{\raise 0.511ex
    \hbox{$<$}}{\lower 0.511ex \hbox{$\sim$}}}}

\renewcommand{\vec}[1]{\boldsymbol{#1}}

\DeclareMathAccent{\ring}{\mathalpha}{operators}{"17}

\providecommand*{\degree}{\ensuremath{^\circ}}

\providecommand{\renewoperator}[3]{\renewcommand*{#1}{\mathop{#2}#3}}
\renewoperator{\Re}{\mathrm{Re}}{\nolimits}
\renewoperator{\Im}{\mathrm{Im}}{\nolimits}

\makeatletter
\providecommand*{\diff}{\@ifnextchar^{\DIfF}{\DIfF^{}}}
\def\DIfF^#1{\mathop{\mathrm{\mathstrut d}}\nolimits^{#1}\gobblespace}
\def\gobblespace{\futurelet\diffarg\opspace}
\def\opspace{%
    \let\DiffSpace\!%
    \ifx\diffarg(%
        \let\DiffSpace\relax
    \else
        \ifx\diffarg[%
            \let\DiffSpace\relax
        \else
            \ifx\diffarg\{%
                \let\DiffSpace\relax
            \fi\fi\fi\DiffSpace}

\newcommand{\II}{$ \mathrm{I\hspace{-.1em}I}\ $}

\shorttitle{DISK STRUCTURE AROUND L1489 IRS}
\shortauthors{Sai et al.}
\begin{document}
%
% ------------------- Title -----------------------------------
\title{DISK STRUCTURE AROUND THE CLASS I PROTOSTAR L1489 IRS REVEALED BY ALMA:\\ A WARPED DISK SYSTEM}
%
% --------------------- Authors ---------------------------------
\author{Jinshi Sai}
\affiliation{Department of Astronomy, Graduate School of Science, The University of Tokyo, 7-3-1 Hongo, Bunkyo-ku, Tokyo 113-0033, Japan}
\affiliation{Subaru Telescope, National Astronomical Observatory of Japan, 650 North A'ohoku Place, Hilo, HI 96720, USA}
\affiliation{Academia Sinica Institute of Astronomy and Astrophysics, 11F of Astro-Math Bldg, 1, Sec. 4, Roosevelt Rd, Taipei 10617, Taiwan}
\author{Nagayoshi Ohashi}
\affiliation{Subaru Telescope, National Astronomical Observatory of Japan, 650 North A'ohoku Place, Hilo, HI 96720, USA}
\affiliation{Academia Sinica Institute of Astronomy and Astrophysics, 11F of Astro-Math Bldg, 1, Sec. 4, Roosevelt Rd, Taipei 10617, Taiwan}
\author{Kazuya Saigo}
\affiliation{Chile Observatory, National Astronomical Observatory of Japan, Osawa 2-21-1, Mitaka, Tokyo 181-8588, Japan}
\author{Tomoaki Matsumoto}
\affiliation{Faculty of Humanity and Environment, Hosei University, Fujimi, Chiyoda-ku, Tokyo 102-8160, Japan}
\author{Yusuke Aso}
\affiliation{Academia Sinica Institute of Astronomy and Astrophysics, 11F of Astro-Math Bldg, 1, Sec. 4, Roosevelt Rd, Taipei 10617, Taiwan}
\affil{Korea Astronomy and Space Science Institute (KASI), 776 Daedeokdae-ro, Yuseong-gu, Daejeon 34055, Republic of Korea}
\author{Shigehisa Takakuwa}
\affiliation{Academia Sinica Institute of Astronomy and Astrophysics, 11F of Astro-Math Bldg, 1, Sec. 4, Roosevelt Rd, Taipei 10617, Taiwan}
\affiliation{Department of Physics and Astronomy, Graduate School of Science and Engineering, Kagoshima University, 1-21-35 Korimoto, Kagoshima, Kagoshima 890-0065, Japan}
\author{Yuri Aikawa}
\affiliation{Department of Astronomy, Graduate School of Science, The University of Tokyo, 7-3-1 Hongo, Bunkyo-ku, Tokyo 113-0033, Japan}
\author{Ippei Kurose}
\affiliation{Department of Astronomy, Graduate School of Science, The University of Tokyo, 7-3-1 Hongo, Bunkyo-ku, Tokyo 113-0033, Japan}
\author{Hsi-Wei Yen}
\affiliation{Academia Sinica Institute of Astronomy and Astrophysics, 11F of Astro-Math Bldg, 1, Sec. 4, Roosevelt Rd, Taipei 10617, Taiwan}
\author{Kohji Tomisaka}
\affiliation{National Astronomical Observatory of Japan, Osawa, 2-21-1, Mitaka, Tokyo 181-8588, Japan}
\author{Kengo Tomida}
\affiliation{Department of Earth and Space Science, Osaka University, Toyonaka, Osaka 560-0043, Japan}
\author{Masahiro N. Machida}
\affiliation{Department of Earth and Planetary Sciences, Faculty of Sciences, Kyushu University, Fukuoka, Fukuoka 819-0395, Japan}
%
% ------------------------------ Abstract----------------------------
\begin{abstract}
We have observed the Class I protostar L1489 IRS with the Atacama Millimeter/submillimeter Array (ALMA) in Band 6. The \ce{C^18O} $J=$2--1 line emission shows flattened and non-axisymmetric structures in the same direction as its velocity gradient due to rotation. We discovered that the \ce{C^18O} emission shows dips at a radius of $\sim$200--300 au while the 1.3 mm continuum emission extends smoothly up to $r\sim400$ au. At the radius of the \ce{C^18O} dips, the rotational axis of the outer portion appears to be tilted by $\sim$15\degree from that of the inner component. Both the inner and outer components with respect to the \ce{C^18O} dips exhibit the $r^{-0.5}$ Keplerian rotation profiles until $r\sim600$ au. These results not only indicate that a Keplerian disk extends up to $\sim$600 au but also that the disk is warped. We constructed a three dimensional warped disk model rotating at the Keplerian velocity, and demonstrated that the warped disk model reproduces main observed features in the velocity channel maps and the PV diagrams. Such a warped disk system can form by mass accretion from a misaligned envelope. We also discuss a possible disk evolution scenario based on comparisons of disk radii and masses between Class I and Class \II sources.
\end{abstract} 
% --------------------------------------------------------------------
%
% -------------- SECTION 1: INTRODUCTION -------------------
\section{Introduction\label{introduction}}
% Large and basic story
Protoplanetary disks are expected to form universally during the star formation process \citep{Terebey:1984aa,Shu:1987aa}. They play key roles as mass reservoirs to central stars in early evolutionary stages and as sites of planet formation in later evolutionary stages.
% disks around T Tauri and recent suggestion about planet formation
These disks have been ubiquitously kinematically confirmed around T Tauri stars (or Class II sources) based on line observations. \citep{Guilloteau:1998aa,Simon:2000aa}.
% Class 0 and I stages
Protostars (or Class 0 and I sources) are considered to be in the disk forming phase because envelopes around protostars provide mass to disks. In fact, previous observations have unveiled disk-like structures around protostars that show rotation and infall velocity features \citep{Ohashi:1997aa,Jorgensen:2009aa}.
% Observations toward Class 0 and I disks
In the past decade, Keplerian disks have been reported for around tens of protostars \citep{Lee:2010aa, Takakuwa:2012aa,Yen:2013aa,Harsono:2014aa}. ALMA observations at high angular resolutions allow us to investigate velocity structures in more detail and determine disk radii through the kinematical transition from envelopes to disks \citep{Ohashi:2014aa,Aso:2015aa,Aso:2017ab}. Such observations suggest that Keplerian disks with radii of 100 au likely form during the Class 0 stage \citep{Yen:2017aa}.

% substructures in Class 0 or I disks
Recent observations of protostellar systems have shown more detailed and complicated disk and envelope structures. The Class I protostar IRAS 04169+2702 shows a velocity structure, implying that its envelope rotates in the opposite direction to its disk \citep{Takakuwa:2018aa}, which could be explained by the Hall effect, one of non-ideal MHD effects \citep{Tsukamoto:2015ab, Tsukamoto:2017aa}. IRS 43, which is a binary system in the Ophiuchus molecular cloud, has two misaligned circumstellar disks with a circumbinary disk \citep{Brinch:2016aa}. Similar misaligned disks have been found in another binary system, L1551 NE \citep{Takakuwa:2017aa}. A skewed dust structure in the Class 0 protostar L1527 IRS was interpreted as a slightly warped (3\degree--5\degree) disk \citep{Sakai:2019aa}. Some theoretical works that carefully consider magnetic fields or turbulence in fact suggest that such misaligned or warped disk systems can form \citep{Matsumoto:2017aa,Bate:2018aa,Hirano:2019aa}. Therefore, a more sophisticated disk formation and evolution scenario requires the properties of embedded young disks to be understood in their formation and evolution phases.

% about L1489 IRS
In this paper, we report observational results of the Class I protostar L1489 IRS in the Taurus molecular cloud, which is one of the nearest star-forming regions to our solar system ($d\sim$140 pc), at a high angular resolution of $\sim \ang[angle-symbol-over-decimal]{;;0.3}$ (corresponding to $\sim$40 au). The bolometric luminosity and bolometric temperature of L1489 IRS are 3.5 $L_\odot$ and 226 K, respectively \citep{Green:2013aa}. This source is surrounded by an envelope having a radius of $\sim$2000 au and mass of $\sim$0.02--0.03$\Msun$ \citep{Motte:2001aa,Hogerheijde:2000aa}; it is also associated with a bipolar outflow ejected in the northern and southern directions at a scale of thousands of au \citep{Tamura:1991aa,Hogerheijde:1998aa,Yen:2014aa}. Therefore, L1489 IRS is still in an accretion phase, although this is well evolved source according to its high bolometric temperature. In fact, infall and rotational motions in the envelope were detected on $\sim$2000 au scales \citep{Hogerheijde:2001aa}. The presence of a large Keplerian disk reaching 700 au in radius with a stellar mass of $\sim$1.6$\Msun$ has been reported based on previous observations with SMA and ALMA at angular resolutions of $\gtrsim \ang[angle-symbol-over-decimal]{;;1}$ \citep{Brinch:2007aa,Yen:2013aa,Yen:2014aa}. However, the internal structures of the disk are still unclear due to the coarse angular resolutions used.

To reveal the disk structures in more detail at a scale of tens of au, we have conducted observations at angular resolutions three times higher than the previous observations \citep{Yen:2014aa} with ALMA in \ce{CO} $J=$2--1 isotopologue lines and 1.3 mm continuum emission. 

The outline of this paper is as follows. Our observations and data reduction are described in Section \ref{observations}. In Section \ref{results}, our observational results are presented. We analyze the morphology and velocity structures of the \ce{C^18O} $J=$2--1 emission in Section \ref{analysis}. In Section \ref{discussion}, we discuss the relation between our findings and previous work, the origin of the suggested disk structures, and disk evolution. Finally, we summarize our results and discussions in Section \ref{summary}.
% --------------------------------------------------------------------
%
% ------- SECTION 2: OBSERVATIONS ---------------------------
\section{Observations\label{observations}}
% ############## table ##############
\begin{table*}[thbp]
\begin{threeparttable}
	\begingroup
	\scalefont{0.9}
	\centering
	\caption{Summary of ALMA Band 6 Observational Parameters}
	\label{obs_summary}
	\begin{tabular*}{2\columnwidth}{@{\extracolsep{\fill}} lccc}
	\hline \hline
	 & \ce{^13CO} $ J=$2--1 & \ce{C^18O} $J=$2--1 & 1.3 mm continuum \\
	\hline
	Date & \multicolumn{3}{c}{2015.May.24 / 2015.Sep.20 \tnote{a}} \\
	Target & \multicolumn{3}{c}{L1489 IRS} \\
	Phase center &  \multicolumn{3}{c}{R.A. (J2000) = $4^\mathrm{h}4^\mathrm{m}42.85^\mathrm{s}$} \\
	 & \multicolumn{3}{c}{Dec. (J2000) = $+26\degree18'56''.3$} \\
	Primary beam & \multicolumn{3}{c}{$\ang[angle-symbol-over-decimal]{;;28.6}$} \\
	Baseline & \multicolumn{3}{c}{ 20.6 m--558.2 m / 40.6 m--1507.9 m \tnote{a}} \\
	Frequency & 220.39868 GHz & 219.56035 GHz & 234.000 GHz \\
	Synthesized beam (P.A.) & $\ang[angle-symbol-over-decimal]{;;0.35} \times \ang[angle-symbol-over-decimal]{;;0.25}$ ($\ang{30}$) & $\ang[angle-symbol-over-decimal]{;;0.33} \times \ang[angle-symbol-over-decimal]{;;0.24}\ (\ang{30})$ & $ \ang[angle-symbol-over-decimal]{;;0.34} \times \ang[angle-symbol-over-decimal]{;;0.23}\ (\ang{28})$ \\
	Velocity resolution (lines) / Band width (continuum) & $0.17 \kmps$ & $0.17 \kmps$ & 2 GHz \\
	Noise level & $5.9\mjpbm$ & $5.1\mjpbm$ & $ 0.079 \mjpbm$ \\
	\multicolumn{1}{l}{Bandpass calibrator} & \multicolumn{3}{c}{J0423-0120 / J0510+1800 \tnote{a}} \\
	\multicolumn{1}{l}{Flux calibrator} &\multicolumn{3}{c}{J0510+180 / J0423-013\tnote{a}} \\
	\multicolumn{1}{l}{Phase calibrator} & \multicolumn{3}{c}{J0510+1800 / J0429+2724 \tnote{a}} \\
	\hline
	\end{tabular*}
	% footnote of table
	\begin{tablenotes}
	\item[a] compact configuration/extend configuration
	\end{tablenotes}
	\endgroup
\end{threeparttable}
\end{table*}
% #####################################
% contents
We have observed L1489 IRS with ALMA in Band 6 during its Cycle 2 phase. The observations were conducted with two configurations: one at the baseline length from 20.6 m to 558.2 m on 2015 May 24 and a more extended one at the baseline length from 40.6 m to 1507.9 m on 2015 September 20. The shortest projected baseline, $\sim$13 m with elevation $\sim$40\degree, is comparable with that in previous ALMA observations \citep{Yen:2014aa}, which provides sensitivity to structures extending to $\sim \ang[angle-symbol-over-decimal]{;;17}$ scales at a level of 10\% \citep{Wilner:1994aa}. The phase center during the two tracks was $\alpha (\mathrm{J}2000)=4^\mathrm{h}4^\mathrm{m}42^\mathrm{s}.85$, $\delta (\mathrm{J}2000)=+\ang[angle-symbol-over-decimal]{26;18;56.30}$. The on-source times are $\sim$23 min and $\sim$25 min for the former and latter observations, respectively. We have observed the 1.3 mm continuum, the \ce{^13CO} $J=$2--1 line emission, and \ce{C^18O} $J=$2--1 line emission. All details of our observations are provided in Table \ref{obs_summary}.

The calibration of the data with longer and shorter baselines was done with CASA 4.5.0 and CASA 4.5.2, respectively. The flux scales are calibrated with J0510+180 and J0423-013, and have an uncertainty of $\sim$10\% \citep{Lundgren:2013aa}. All of our images were produced by CLEAN using the CASA task \textit{clean} with CASA version 5. For all the images, the Briggs weighting with a robust vale of 0.5 was used. In the CLEAN process, the multiscale option to recover extended emission was used with scales sizes of approximately zero, three, five and ten times as large as the synthesized beam size (a point source, $\ang[angle-symbol-over-decimal]{;;0.75}$, $ \ang[angle-symbol-over-decimal]{;;1.25}$, and $\ang[angle-symbol-over-decimal]{;;2.5}$). The synthesized beam sizes are $\ang[angle-symbol-over-decimal]{;;0.35} \times \ang[angle-symbol-over-decimal]{;;0.25}$ ($\ang{30}$) in the \ce{^13CO} $J=$2--1 line emission, $\ang[angle-symbol-over-decimal]{;;0.33} \times \ang[angle-symbol-over-decimal]{;;0.24}\ (\ang{30})$ in the \ce{C^18O} $J=$2--1 line emission, and $\ang[angle-symbol-over-decimal]{;;0.34} \times \ang[angle-symbol-over-decimal]{;;0.23}\ (\ang{28})$ in the 1.3 mm continuum. The root mean square (rms) noise levels, which do not include the systemic uncertainty of the flux calibration, are 5.9$\mjpbm$ in the \ce{^13CO} $J=$2--1 line emission and 5.1$\mjpbm$ in the \ce{C^18O} $J=$2--1 line emission. All velocity resolutions are 0.17$\kmps$. We also performed self-calibration on the 1.3 mm continuum data using CASA tasks \textit{clean}, \textit{gaincal}, and \textit{applycal} to improve the signal-to-noise ratio (SNR) of the continuum image. Only the phase was calibrated with a solution interval of each scan (i.e., with the time bin of $\sim$6 s). The number of iterations was one, which increased the peak SNR from $\sim$59 to $\sim$97. The phase-only self-calibration also increased the total flux density by $\sim$20\%. The final rms noise level for the 1.3 mm continuum is 0.079$\mjpbm$. In this paper, we mostly focus on the observational results for the 1.3 mm continuum and the \ce{C^18O} $J=$2--1 line because the \ce{^13CO} emission shows similar results to those obtained in the \ce{C^18O} emission. The results of the \ce{^13CO} $J=$2--1 line are presented in Appendix \ref{app_13co}.
% -----------------------------------------------------
%
% ------------------ SECTION 3: Results -----------------------
\section{Results\label{results}}
% Continuum
\subsection{1.3 mm Continuum Emission}
% ###########  Figure: 1.3 mm continuum #####
\begin{figure}
\centering
\includegraphics[viewport=130 20 712 575, width=\the\columnwidth]{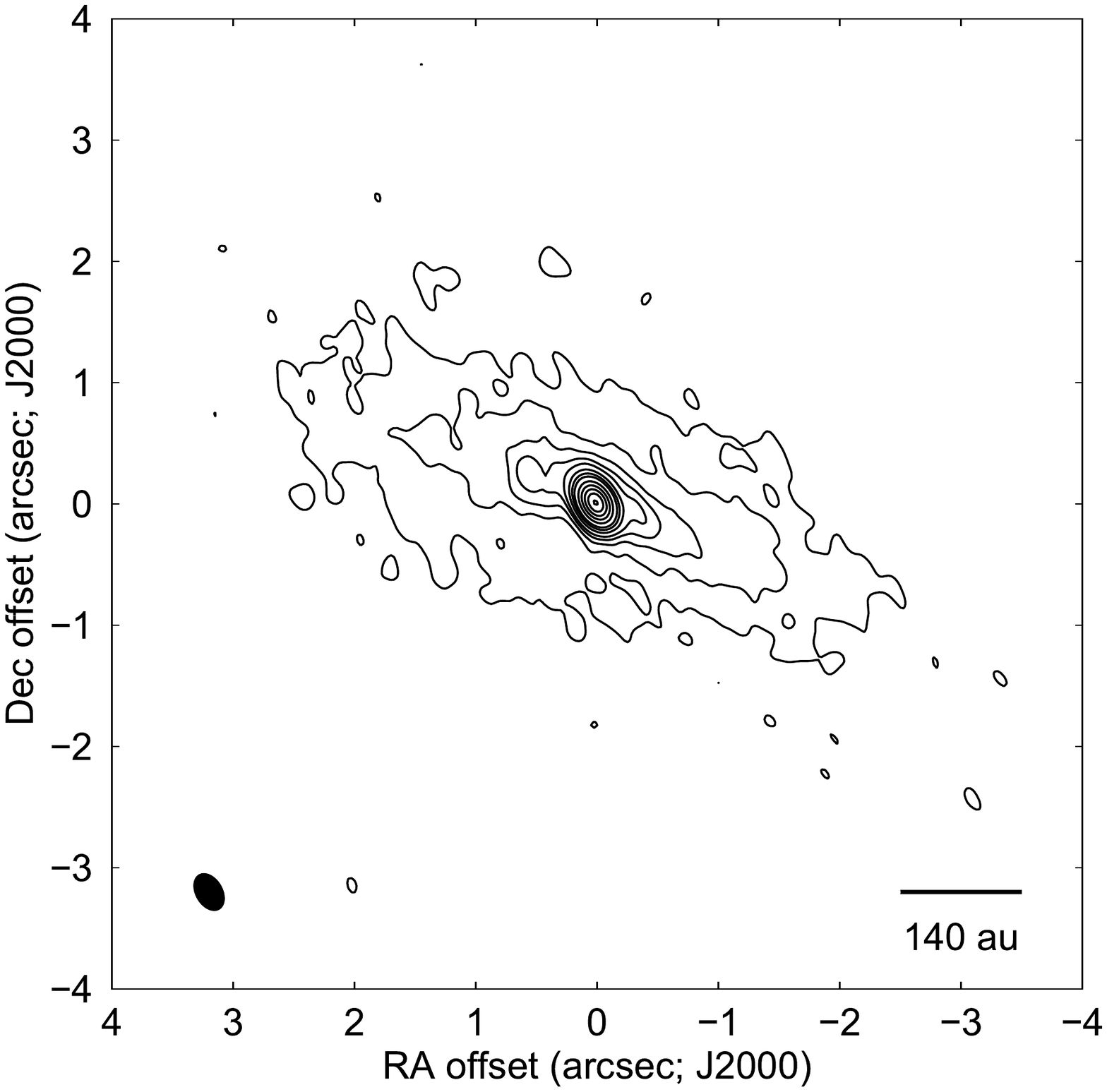} % 0 0 842 595
\caption{1.3 mm continuum emission map of L1489 IRS. Contour levels are from 3$\sigma$ to 15$\sigma$ in steps of 3$\sigma$, from 15$\sigma$ to 30$\sigma$  in steps of 5$\sigma$,  and from 30$\sigma$ to 90$\sigma$ in steps of 10$\sigma$, where 1$\sigma$ corresponds to $0.079 \mjpbm$. The black-filled ellipse in the bottom-left corner indicates the synthesized beam size: $\ang[angle-symbol-over-decimal]{;;0.34}\times \ang[angle-symbol-over-decimal]{;;0.23}$, P.A.$=\ang{28}$.}
\label{1.3mmcont}
\end{figure}
% ####################################
% contents
Figure \ref{1.3mmcont} shows the self-calibrated 1.3 mm continuum image of L1489 IRS. The overall structure is elongated from NE to SW. A similar elongated structure of the 1.3 mm continuum emission was obtained in previous observations \citep{Yen:2014aa}; however, the present angular resolution is almost three times higher than the previous one, which allows us to investigate the spatial structures in more detail. The compact emission around the center is resolved in the present observations. The peak position measured before self-calibration is at $\alpha (\mathrm{J}2000)=4^\mathrm{h}4^\mathrm{m}43^\mathrm{s}.07$, $\delta (\mathrm{J}2000)=+\ang[angle-symbol-over-decimal]{26;18;56.2}$, which is consistent with that in previous observations \citep{Yen:2014aa}. In this paper, we treat this position as the position of the central protostar, and all maps are centered on this position.

The observed 1.3 mm continuum is well fitted by two Gaussian components---a central compact component and a broad component---with residuals at most $5.5\sigma$ but mostly less than 3$\sigma$, as was also highlighted by \cite{Yen:2014aa}. The fitting was performed with the MIRIAD fitting task \textit{imfit}. The deconvolved size of the central compact component is $\ang[angle-symbol-over-decimal]{;;0.097} \times \ang[angle-symbol-over-decimal]{;;0.037}$ ($\sim$14 au$\times 5$ au), and the position angle is $49\degree$. The total flux density of the central compact component is $5.9 \pm 0.32$ mJy, which is slightly less than the $7.6 \pm 0.5$ mJy measured by \cite{Yen:2014aa}. On the other hand, the total flux density of the broad component is $\sim$53$\pm 4.5$ mJy, which is larger than the $42 \pm 3.7$ mJy measured by \cite{Yen:2014aa}. This would be because the central compact component was spatially unresolved in the previous observations. The total flux density derived from the two Gaussian components is 59 $\pm$ 4.5 mJy, which is $\sim10$ mJy larger than the value measured in \cite{Yen:2014aa}. This increase of the total flux density is greater than the uncertainty of the flux calibration of $\sim$10\%, and likely due to the self-calibration as mentioned in Section \ref{observations}. The deconvolved size of the broad Gaussian component is $ \ang[angle-symbol-over-decimal]{;;4.1} \times \ang[angle-symbol-over-decimal]{;;1.2}$ (corresponding to $\sim$570 au$\times 170$ au), and the position angle is $69\degree$, which are both consistent with previous observations \citep{Yen:2014aa}. Assuming that the 1.3 mm continuum emission arises from a geometrically thin disk, we estimate the inclination angle $i$ of the disk to be $\sim$73\degree from the lengths of the major and minor axes of the broad Gaussian component ($i=90 \degree$ corresponds to the edge-on configuration).

% disk mass
The disk gas mass is estimated from the total continuum flux density with the following equation:
 \begin{align}
 M_\mathrm{disk}=\frac{F_\mathrm{\nu}d^2}{\kappa_\mathrm{\nu}B_\mathrm{\nu}(T_\mathrm{disk})},
 \label{equation_diskmass}
 \end{align}
 where $F_\mathrm{\nu}$ is the total flux density of the continuum emission, $d$ is the distance to the object, $\kappa_\mathrm{\nu}$ is the dust mass opacity at frequency $\nu$, $B_\nu$ is the Plank function, and $T_\mathrm{disk}$ is the mean disk temperature. We assumed the distance to be 140 pc, and the dust temperature to be 25 K, which is the minimum disk temperature within a radius of $500$ au in a disk model in \cite{Yen:2014aa}. This dust temperature is consistent with the temperature suggested from our calculations of a disk model with RADMC-3D, and that suggested in a previous work \citep[see Section \ref{origin_of_gap} for more information]{Brinch:2007aa}. The dust mass opacity is assumed to follow $\kappa_\mathrm{\nu} = 0.1 (\nu/1 \mathrm{THz})^\beta$ with $\beta = 1$, resulting in $\kappa_\mathrm{\nu} = 0.023\ \mathrm{g\ cm^{-2}}$ at 234 GHz \citep{Beckwith:1990aa,Jorgensen:2007aa}.
 The total disk gas mass is thus estimated to be $7.1 \times 10^{-3} \Msun$ with $F_\nu = 59$ mJy. When using the dust opacity $\sim$0.90 $\mathrm{g\ cm^{-2}}$ derived by \cite{Ossenkopf:1994aa}, a total disk gas mass of $1.8 \times 10^{-2} \Msun$ is obtained.
% C18O
% ########## Figures: C18O ############
% moment maps
\begin{figure}
\centering
\includegraphics[viewport=120 30 722 565, width=\the\columnwidth]{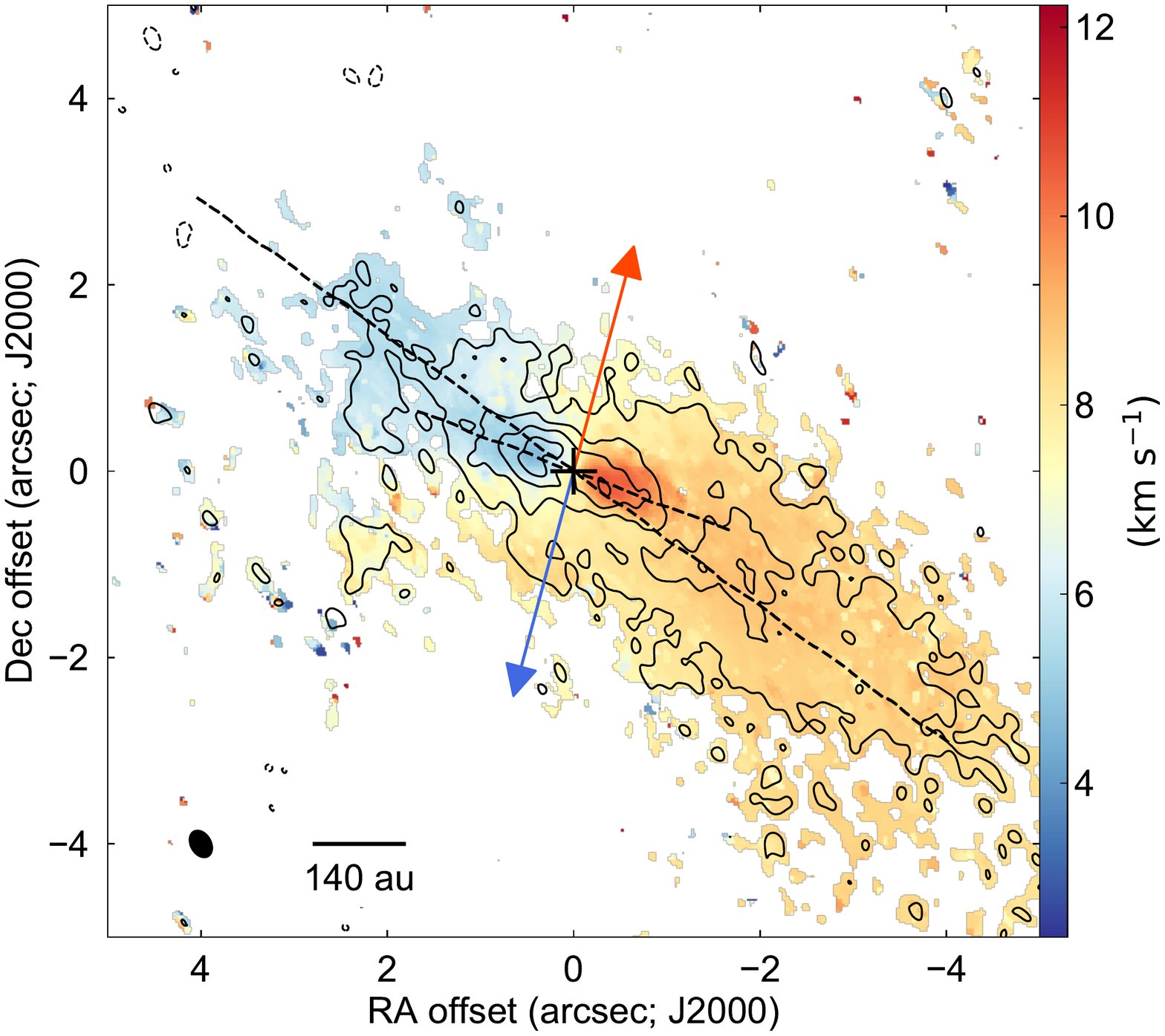} % 0 0 842 595, rotated
\caption{Total intensity map (\textit{contours}) and mean velocity map (\textit{color}) of the \ce{C^18O} $J=$2--1 emission of L1489 IRS. Contour levels are -3, 3, 6, 9, 12 $\times \sigma$, where 1$\sigma$ corresponds to $8.2 \mjpbm \kmps$. Negative contours are shown with dashed lines. The cross shows the position of the central protostar (i.e., 1.3 mm continuum emission peak). The blue and red arrows denote the direction of the blueshifted and redshifted components of the outflow, respectively, with a position angle of $165\degree$ \citep{Hogerheijde:1998aa}. The black-filled ellipse in the bottom-left corner indicates the synthesized beam size: $\ang[angle-symbol-over-decimal]{;;0.33}\times \ang[angle-symbol-over-decimal]{;;0.24}$, P.A.$=\ang{30}$. Dashed lines show the direction of elongation at a small scale of $r\sim140$ au and a large scale of $r\sim600$ au.}
\label{momc18o21}
\end{figure}
%
% 1D profile
\begin{figure}
\centering
\includegraphics[viewport=20 30 542 565, width=\the\columnwidth]{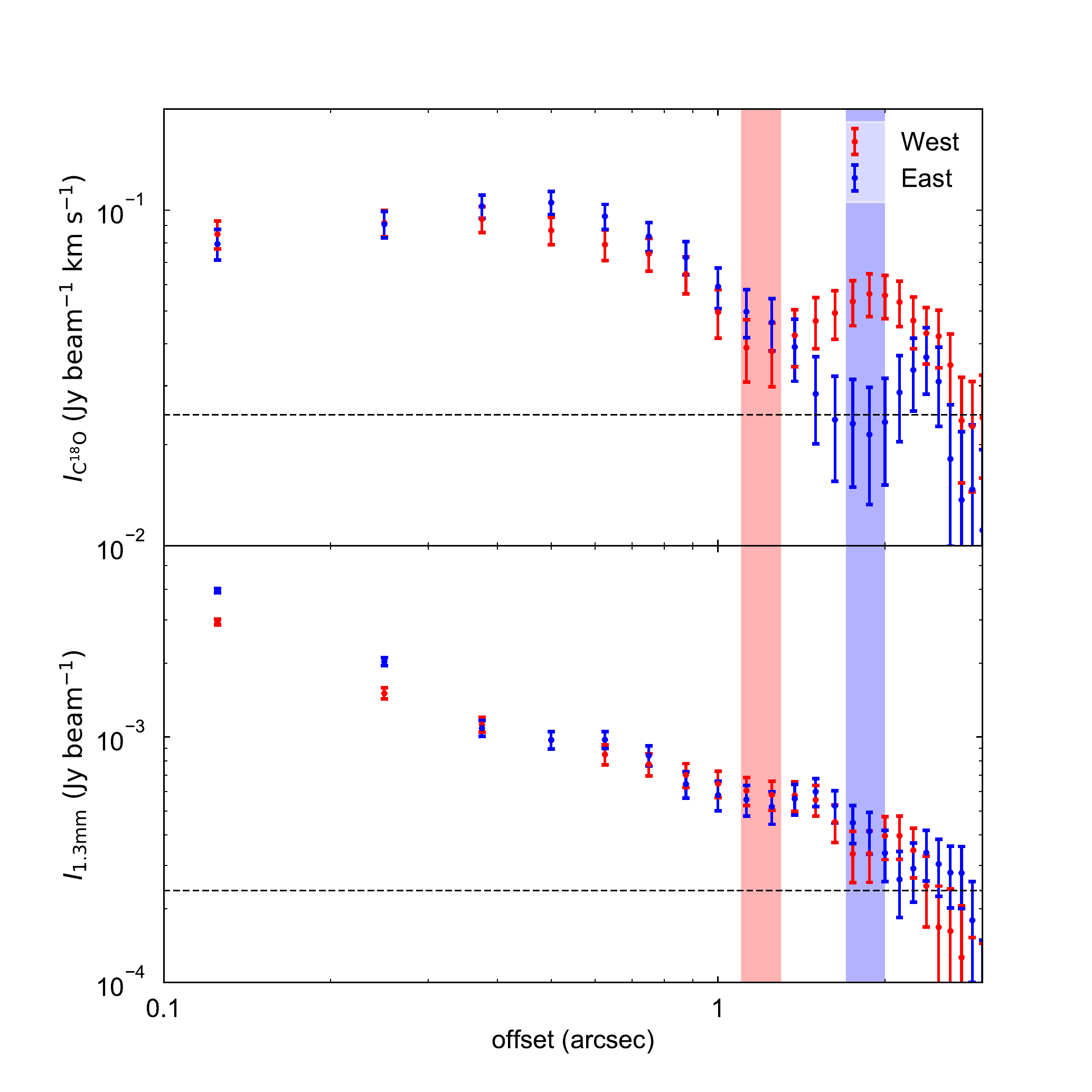} % viewport: bottom right top left (cuz it was rotated)
\caption{Radial intensity profiles of the integrated \ce{C^18O} $J=$2--1 emission (\textit{top}) and 1.3 mm continuum (\textit{bottom}) along the \ce{C^18O} elongated structure on a small scale (P.A.$= 69\degree$). Blue and red colors show the profiles of the northeastern blueshifted and southwestern redshifted emissions, respectively. The error bars show the rms of each image. Filled regions denote the gap positions seen in the \ce{C^18O} emission. Dashed lines show 3$\sigma$ noise levels.}
\label{1dprof}
\end{figure}
%
% channel map
\begin{figure*}
\centering
\includegraphics[width=2\columnwidth]{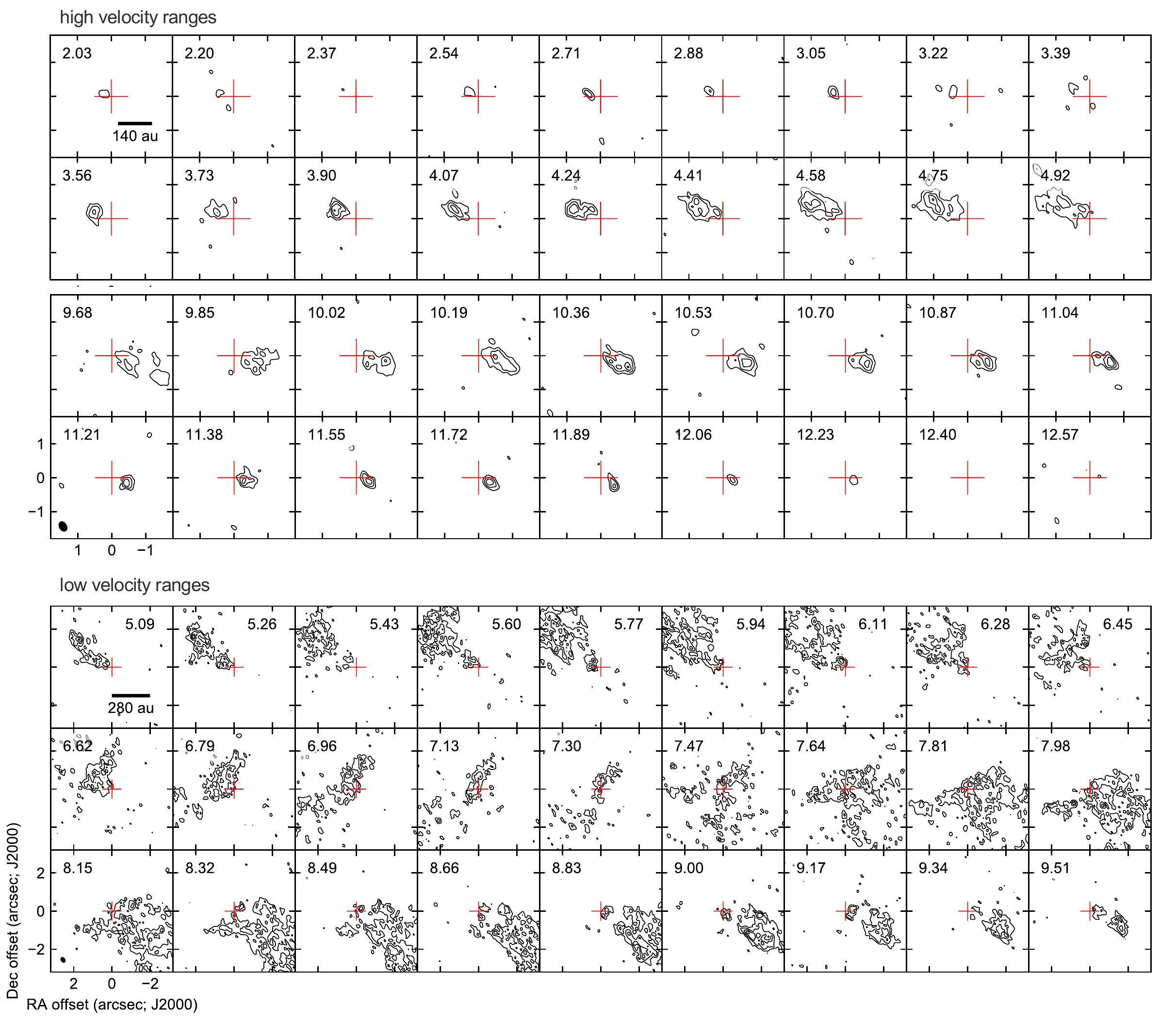}
\caption{Velocity channel maps of the \ce{C^18O} $J=$2--1 emission at the higher (\textit{upper panels}) and lower velocity ranges (\textit{lower panels}). Note that the spatial coverage is distinct between the two panels in order to show the compact high-velocity emission more clearly. Contour levels are 3, 5, 7, 9, ... $\times \sigma$, where 1$\sigma$ corresponds to $5.1 \mjpbm$. The LSRK velocity of each panel is indicated in the top-left (or top-right) corner in $\kmps$. The systemic velocity is 7.3$\kmps$. The red cross at the center of each panel shows the position of the central protostar (i.e., the 1.3 mm continuum emission peak). The black-filled ellipses at the bottom-left corners denote the synthesized beam size: $\ang[angle-symbol-over-decimal]{;;0.33}\times \ang[angle-symbol-over-decimal]{;;0.24}$, P.A.$=\ang{30}$.}
\label{channelc18o21}
\end{figure*}
%
% moment 0 map over different velocity range
\begin{figure*}[thbp]
\centering
\includegraphics[width=2\columnwidth]{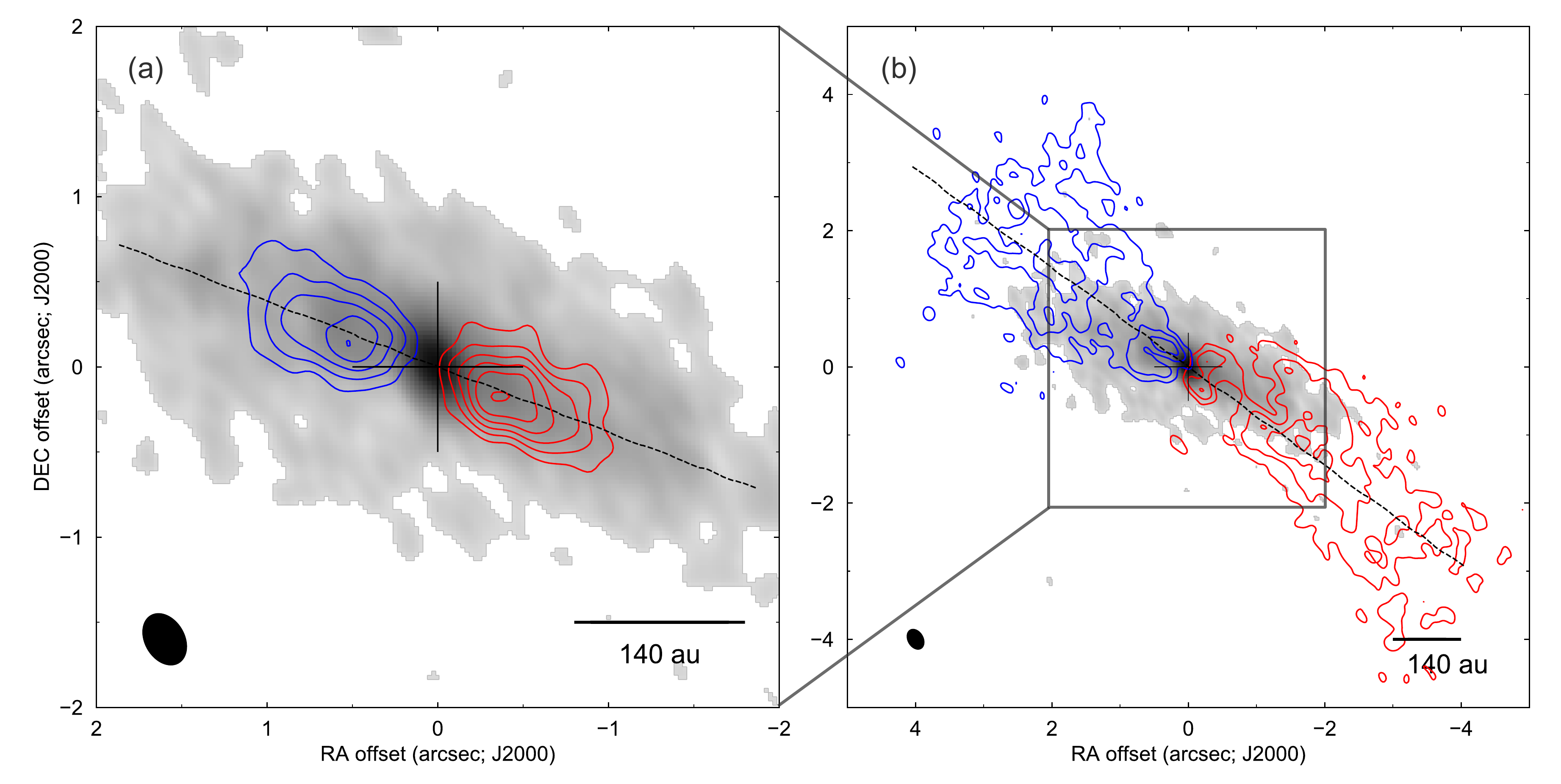}
\caption{Integrated intensity maps of (a) the high velocity components integrated over $2.55 \kmps \le | \vsys - \vlsr| \le 4.59 \kmps$ and (b) the low velocity components integrated over $1.19 \kmps \le | \vsys - \vlsr| \le 2.21 \kmps$ of the \ce{C^18O} $J=$2--1 emission of L1489 IRS (\textit{color contours}) overlaid on the 1.3 mm continuum map above $3\sigma$ (\textit{gray scale}).
Blue and red contours show the integrated emissions of the blueshifted and redshifted components, respectively, with contour levels of 3, 6, 9, 12, ...$\times \sigma$. For the high-velocity component, 1$\sigma$ corresponds to $5.0 \mjpbm \kmps$ for the blueshifted emission and $4.9 \mjpbm \kmps$ for the redshifted emission. For the low-velocity component, 1$\sigma$ corresponds to $3.6 \mjpbm \kmps$ for the blue shifted emission and $3.9 \mjpbm \kmps$ for the redshifted emission. Black crosses show the position of the central protostar (i.e., 1.3 mm continuum emission peak). Black-filled ellipses in the bottom-left corners indicate the observational beam size: $\ang[angle-symbol-over-decimal]{;;0.33}\times \ang[angle-symbol-over-decimal]{;;0.24}$, P.A.$=\ang{30}$. Dashed lines denote the position angle of the velocity gradient of each component.}
\label{lowhighvelmom}
\end{figure*}
% ########################################
% overall description
\subsection{\ce{C^18O} $J=$2--1 \label{result_c18o}}
Figure \ref{momc18o21} shows the total intensity map (i.e., moment 0 map) integrated over the velocity from 2.37 $\kmps$ to 12.23 $\kmps$ and the mean velocity map (i.e., moment 1 map) of the \ce{C^18O} $J=$2--1 line emission of L1489 IRS. The moment 0 and 1 maps of the \ce{^13CO} line are shown in Figure \ref{mom13co21} (see Appendix \ref{app_13co}). The overall structure of the \ce{C^18O} total intensity map is flattened and elongated from NE to SW, similar to that of the continuum emission. Its elongation is almost perpendicular to the molecular outflow direction \citep{Tamura:1991aa,Hogerheijde:1998aa,Yen:2014aa}. The overall velocity structure is described as blueshifted velocities on the NE side of the central protostar and redshifted velocities on the other side with respect to the systemic velocity of $7.3\kmps$ of the L1489 IRS system. This velocity gradient from NE to SW in the direction of the elongation is interpreted as rotation around the protostar. The emission tail to the southwest is much longer than its northeastern counterpart, showing a non-axisymmetric morphology with respect to the central protostellar position. The elongation is $\sim$1100 au across and broader than that of the 1.3 mm continuum. The total intensity map shows a double peak centered around the protostellar position, which was not spatially resolved in previous ALMA observations \citep{Yen:2014aa}, whereas the continuum exhibits a single peak.

% about gap in 1d profile
A further careful inspection of the integrated intensity map reveals that the \ce{C^18O} emission becomes weaker at $\sim \ang[angle-symbol-over-decimal]{;;2}$ NE and at $\sim \ang[angle-symbol-over-decimal]{;;1}$ SW of the central star. To investigate these dips in the \ce{C^18O} emission, the radial intensity profile of the \ce{C^18O} integrated emission along the direction of the \ce{C^18O} elongated structure on a small scale ($69\degree$; see below), is presented in Figure \ref{1dprof}. The intensity profile more clearly exhibits the dips of the \ce{C^18O} emission at positions marked by the zones filled in blue and red. For comparison, a radial intensity profile of the 1.3 mm continuum emission along the same direction is shown in the bottom panel of Figure \ref{1dprof}. Interestingly, the 1.3 mm continuum emission does not show such dips at the positions of the \ce{C^18O} dips, suggesting that these \ce{C^18O} dips are not due to physical absence of material. The origin of the \ce{C^18O} dips are discussed in more detail in Section \ref{origin_of_gap}.

% about channel maps
More details of the velocity structures of the \ce{C^18O} emission can be seen in the velocity channel maps presented in Figure \ref{channelc18o21}. The \ce{^13CO} $J=$2--1 channel maps are also shown in Figure \ref{channel13co21} (see Appendix \ref{app_13co}) for comparison. A compact ($\lesssim$140 au) \ce{C^18O} emission is seen near the protostellar position at $\vlsr \le 4.92 \kmps$ and at $\vlsr \ge 9.68 \kmps$ in the upper four rows in Figure \ref{channelc18o21}. These compact components, highly blueshifted and redshifted emission located to the NE and SW sides of the protostar, produce a steep velocity gradient in the close vicinity to the protostar ($r \lesssim \ang[angle-symbol-over-decimal]{;;1}$) from NE to SW, as shown in Figure \ref{momc18o21}. On the other hand, the emission is more spatially extended at the lower velocity of $ 5.09 \kmps < \vlsr \le 9.51 \kmps$, producing a smaller velocity gradient across $\sim1100$ au in Figure \ref{momc18o21}. At the velocity ranges of $ 5.09 \kmps < \vlsr \le 6.28 \kmps$ and $ 8.32 \kmps \le \vlsr < 9.51 \kmps$, the emission is faint at $\sim \ang[angle-symbol-over-decimal]{;;2}$ NE and $\sim \ang[angle-symbol-over-decimal]{;;1}$ SW of the stellar position, respectively. These faint points produce the \ce{C^18O} dips seen in the total intensity map.

% about gap and warped structure
To demonstrate the difference in the \ce{C^18O} morphology between higher and lower velocities more clearly, the moment 0 maps of the high and low velocity components integrated over the velocity ranges of $2.55 \kmps \le | \vsys - \vlsr| \le 4.59 \kmps$ and $1.19 \kmps \le | \vsys - \vlsr| \le 2.21 \kmps$, respectively, are shown in Figure \ref{lowhighvelmom}. As expected, the moment 0 map for the high velocity component shows compact blueshifted and redshifted emissions in close vicinity to the central protostar ($r \lesssim \ang[angle-symbol-over-decimal]{;;1}$), whereas that for the low velocity component shows extended emissions. Velocity gradients between blueshifted and redshifted components are seen in both maps, which are consistent with the moment 1 map shown in Figure \ref{momc18o21}. The dust continuum extends smoothly across the \ce{C^18O} dips as shown in Figure \ref{1dprof}.

Furthermore, it is clear that the directions of the velocity gradients are distinct between the two maps. To quantitatively assess the difference, we fitted 2D Gaussian functions to these blueshifted and redshifted components in the relevant maps with MIRIAD fitting task \textit{imfit} and determined their peak positions. From the line passing through the peaks of the blueshifted and redshifted components, we derived P.A. of the direction of the velocity gradient. Note that the fitting was performed with a single Gaussian function for the high velocity component but with a two-component Gaussian function for the low velocity component to fit the inner peaks within $\ang[angle-symbol-over-decimal]{;;1}$ and outer peaks at  $\ang[angle-symbol-over-decimal]{;;2}$. The position angle of the low velocity component is derived from a line passing through the outer peaks of blueshifted and redshifted components. The P.A. is estimated to be $69\degree \pm 2.6\degree $ for the high velocity component, which is almost the same direction as the dust continuum elongation, while it is estimated to be $54\degree \pm 2.9\degree$ for the low velocity component, showing a difference of $\sim$15\degree between the two components. These velocity ranges of the high and low velocity components are chosen based on the morphology of the emission in the velocity channel map (Figure \ref{channelc18o21}) so that each velocity component reasonably traces rotation motion on different physical scales. This difference in the position angle of the velocity gradients is also seen in Figure \ref{momc18o21}: on a large scale of $r\sim$600 au, the direction of the elongation of the emission is close to P.A.$=54 \degree$, while it approaches P.A.$=69 \degree$ on a small scale within $\sim$140 au in radius from the central star. The velocity gradient changes its position angle from close to $54\degree$ to $69\degree$ along with the elongation. This variation of the position angle of the elongation exhibits a tilted integral shape, which is also seen in previous ALMA observations \citep{Yen:2014aa}.
% --------------------------------------------------------------
%
% ----------------- SECTION 4: Analysis ------------------
\section{Analysis\label{analysis}}
% rotational profile
\subsection{Velocity Structure of the \ce{C^18O} 2-1 Emission\label{analysis_vc18o}}
% ### Figures: PV diagram & rotational profile #####
\begin{figure*}[thbp] % xbb: 0 0 842 595
\centering
\includegraphics[width=1.9\columnwidth]{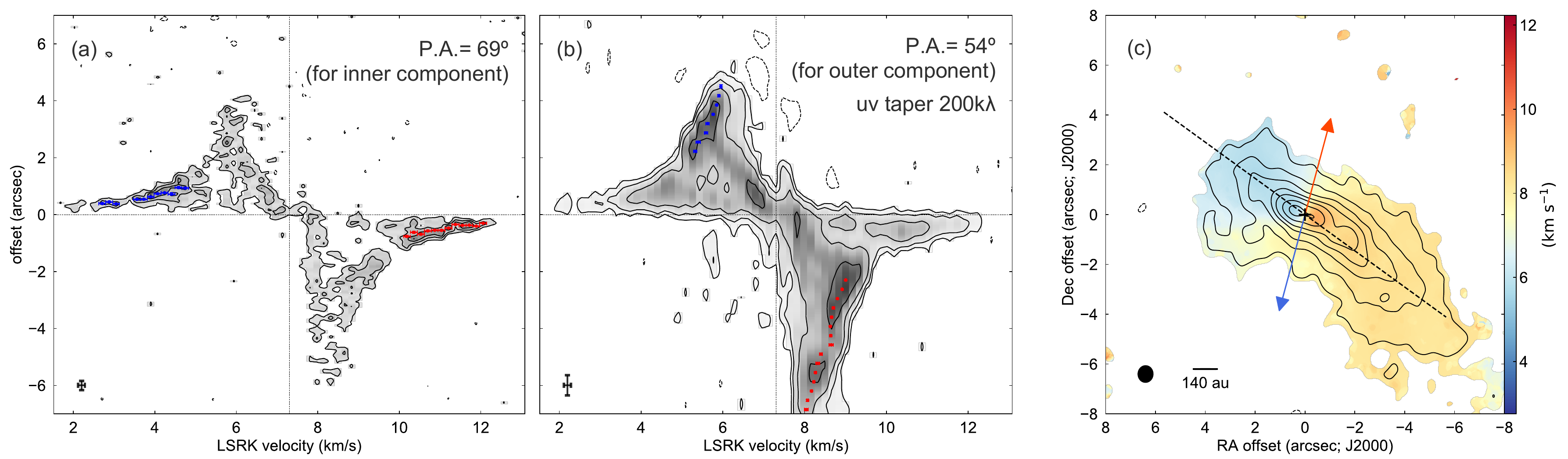}
\caption{(a) Position-velocity (PV) diagram along the velocity gradient of the inner component (P.A.$=69\degree$). Contour levels are from 3$\sigma$ to 7$\sigma$ in steps of 2$\sigma$, where 1$\sigma$ corresponds to $5.1 \mjpbm$. (b) PV diagram produced from an tapered image shown in Figure \ref{pvobs}c along the velocity gradient of the outer component (P.A.$=54\degree$). Contour levels are -3, 3, 6, 12, 24 $\times \sigma$, where 1$\sigma$ corresponds to $5.4 \mjpbm$. Negative contours are shown  with dashed lines. Vertical and horizontal bars at the bottom-left corners denote FWHM of the synthesized beam along its major axis and the velocity resolution, respectively. Blue and red points denote representative velocities and positions also plotted in Figure \ref{rplaw}. (c) Same as Figure \ref{momc18o21} but shows maps made from the visibility tapered at FWHM=$200 \mathrm{k\lambda}$. The synthesized beam size is $\sim \ang[angle-symbol-over-decimal]{;;0.71} \times \ang[angle-symbol-over-decimal]{;;0.64}$, P.A.$=\ang{-1.9}$. Contour levels are from $3\sigma$ to $15\sigma$ in steps of 3$\sigma$, and from 15$\sigma$ to 25$\sigma$ in steps of $5\sigma$, where $1\sigma$ corresponds to $16\mjpbm \kmps$. A dashed line shows the direction of the axis of the PV diagram shown in Figure \ref{pvobs}b (P.A.$=54\degree$).}
\label{pvobs}
\end{figure*}
% ###########################################
% contents
Figure \ref{pvobs}a and b shows the position-velocity (PV) diagrams along the velocity gradients of the inner ($< \ang[angle-symbol-over-decimal]{;;2}$) high velocity component and the outer ($> \ang[angle-symbol-over-decimal]{;;2}$) low velocity component (along P.A.$=69 \degree$ and P.A.$= 54\degree$, respectively). The PV diagram along P.A.$= 54 \degree$ is produced from the \ce{C^18O} image cube made from the visibility tapered at FWHM = $200\ \mathrm{k\lambda}$ in the $(u,v)$ domain to recover extended structures at an angular resolution enough high for the fitting procedure described later in this section (Figure \ref{pvobs}c). Both PV diagrams show almost the same velocity features. The PV diagrams exhibit so-called spin-up rotation on a scale of up to $r \sim \ang[angle-symbol-over-decimal]{;;4}$--$\ang[angle-symbol-over-decimal]{;;6}$, where the velocity is larger at inner regions than outer regions. The \ce{C^18O} dips are seen at $\sim \ang[angle-symbol-over-decimal]{;;2}$ and $\sim \ang[angle-symbol-over-decimal]{;;1}$ in the blueshifted and redshifted components showing spin-up rotation, respectively. The velocity appears proportional to the offset from the central star over the velocity range of $\sim$6--9$\kmps$. The velocity channel at $\sim$8$\kmps$ shows an extended component over $\ang[angle-symbol-over-decimal]{;;7}$.

To examine the nature of the rotation, we derive rotational profiles from the PV diagrams. We note that similar methods were used in \cite{Yen:2013aa,Ohashi:2014aa,Aso:2015aa,Aso:2017ab}. First, we fit a Gaussian function to the intensity distribution at each velocity channel on the PV diagram cut along P.A.$=69\degree$. Then, we determine a representative position at each velocity for the inner component $\lesssim$100 au. The determined representative points are shown in Figure \ref{pvobs}a. The error bars in the position and velocity axes are the fitting error and the velocity resolution, respectively. In a case where angular resolutions are not sufficient to resolve emission, peak positions of the emission from disks can shift to inner radii due to beam convolution, resulting in incorrect rotational profiles \citep{Aso:2015aa}. To avoid this effect, we use only data points at $r<100 \au$ for fitting to the inner component. Under this condition, where $1/\theta_\mathrm{n}$ in \cite{Aso:2015aa} is calculated as $\sim$0.6 from $R_0 \sim 100$ au, $i \sim 73 \degree$ and $\theta = \ang[angle-symbol-over-decimal]{;;0.33}$, the derived power-law index is correct but the derived central stellar mass should be multiplied by 1.3 to correct the effect of the beam convolution (see APPENDIX A in \cite{Aso:2015aa}). While the correction for the beam size effect described in \cite{Aso:2015aa} is for the intensity weighted mean position, the representative position on the PV diagram determined by Gaussian fit is almost the same as the intensity weighted mean position. Hence, the same correction is adopted in this paper.
% log r- log v plot
% ########## Figure: rotation profile ###########
% rotational power law profile
\begin{figure*}[htbp]
\centering
\includegraphics[width=1.9\columnwidth]{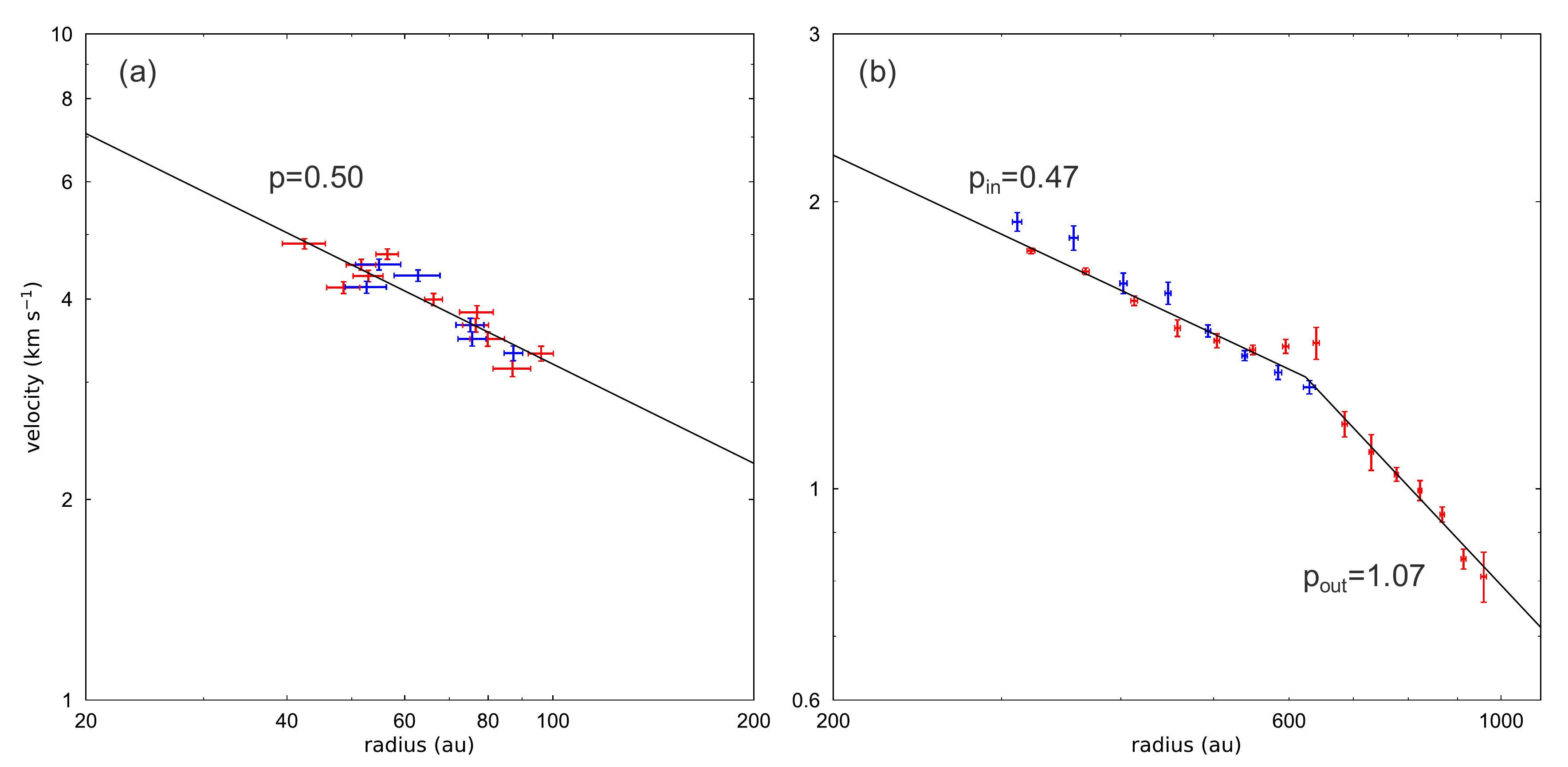}
\caption{Plots of representative points in the PV diagrams cut along the velocity gradients of (a) the inner component (P.A. $= 69\degree$) and (b) the outer component (P.A. $= 54\degree$) on the $\log R$-$\log V$ plane. Blue and red points are the data points of the blueshifted and redshifted components, respectively. The solid lines denote the best fit rotational power-law functions.}
\label{rplaw}
\end{figure*}
% #################################
We fit the following power-law function to the determined data points from the PV diagram by $\chi^2$ fitting:
\begin{align}
\vrot = V_\mathrm{100} \left(
\frac{r}{\mathrm{100} \au}
\right)^{-p},
\label{plaw-sp}
\end{align}
where $\vrot$ is the rotational velocity, $V_\mathrm{100}$ is the rotational velocity at 100 au in radius, $p$ is the power-law index, and $r$ is the radius. $\vsys$ is also treated as a free parameters according to $V_\mathrm{rot} \propto | V_\mathrm{obs} - \vsys |$. The fitting result, along with the measured data points on the PV diagrams, is plotted on the $\log R$-$\log V$ planes without correction for inclination in Figure \ref{rplaw}a. The best fit parameters are summarized in Table \ref{table_plaw_param}.

% ##### Table: the best fit parameters of the pv fitting #####
%
\begin{table}[thbp]
\begin{threeparttable}
    \begingroup
    \centering
    \caption{Parameters of the best-fit power-law functions}
    \begin{tabular*}{\columnwidth}{@{\extracolsep{\fill}}cccc}\hline\hline
    \multicolumn{4}{c}{for the inner component (single power-law)} \\
    $\vsys$ & $ V_{100}$ & $p$ & \\
    $7.22\pm 0.09\ \kmps$ & $ 3.20\pm0.15\ \kmps$ & $0.50\pm 0.11$ & \\
    \hline
    \multicolumn{4}{c}{for the outer component (double power-law)\tnote{a}}\\
    $V_\mathrm{break}$ & $R_\mathrm{break}$ & $p_\mathrm{in}$ &$ p_\mathrm{out}$ \\ 
    $1.31 \pm 0.08\ \kmps$ & $ 624 \pm 62 \au$ & $0.47\pm0.05$ & $1.07\pm0.24$ \\
    \hline
    \end{tabular*}
	% footnote of table
	\begin{tablenotes}
	\item[a] $\vsys$ is fixed to $7.23\ \kmps$.
	\end{tablenotes}
	\endgroup
\end{threeparttable}
\label{table_plaw_param}
\end{table}
The best fit parameters for the inner component are $(\vsys, V_{100}, p) = (7.22 \pm 0.09 \kmps, 3.20 \pm 0.15 \kmps, 0.50 \pm 0.11)$. The uncertainties of the parameters are derived from fitting 10000 times with the data values changed within their errors at each fitting \citep{Yen:2013aa} as $3\sigma$ values of the posterior distribution. For the inner component, a uniform distribution is assumed for the error of velocity because the uncertainty stems from the velocity resolution. A Gaussian distribution is adopted for the position error. The derived rotational power-law index $\sim$0.5 indicates that the high-velocity inner component traces a Keplerian disk. The estimated central stellar mass after the correction is $1.64 \pm 0.12 \Msun$, which comes from the relation of $\vlos = \sqrt{G M_\ast/r}\sin i$ with $i=73 \degree$, $V_{100}=3.20 \kmps$, and $r = 100 \au$. The uncertainty is calculated from the error of $V_{100}$ through the error propagation. The stellar mass can be $\sim$1.57$\Msun$ assuming the perfect edge-on ($i=90 \degree$) configuration. These results are consistent with the central stellar mass $1.6\pm0.5 \Msun$ with $i=66\degree$ within the errors found in previous low-resolution observations by \cite{Yen:2014aa}.

For the outer component $\gtrsim$300 au, a Gaussian function is fitted to the spectrum at each position to determine representative velocities in contrast to the case of the inner component, where the fitting is performed to derive the representative positions. More specifically, the velocities of the emission peaks along the velocity axis of the PV diagram are first derived. The PV diagram along P.A.$=54\degree$ produced from the tapered image shown in Figure \ref{pvobs}c is used for the fitting procedure. Velocity channels in the neighboring $\pm$2 channels of the peak velocities are extracted, and Gaussian fitting to the spectra in the restricted spectral range is performed to derive the centroid velocities. This channel selection method can extract the proper power-law index. Further details of this procedure are given in Appendix \ref{app_pv}. The uncertainties in the position and velocity axes are the position accuracy (angular resolution/SNR) and fitting error, respectively. The determined representative points are shown in Figure \ref{pvobs}b.

Power-law fittings to the determined data points are performed, as in the case of the inner component. In addition to Equation (\ref{plaw-sp}), the following double power-law function is also examined to simultaneously account for the disk and envelope components:
\begin{align}
V_\mathrm{rot} =
	\begin{cases}
	V_\mathrm{break} \left( \frac{r}{R_\mathrm{break}} \right)^{-p_\mathrm{in}}
	 & \left(r \leq R_\mathrm{break} \right) \\
	V_\mathrm{break} \left( \frac{r}{R_\mathrm{break}} \right)^{-p_\mathrm{out}}
	 & \left(r > R_\mathrm{break} \right)
	\end{cases},
\label{plaw-double}
\end{align}
where $R_\mathrm{break}$ is the radius at which the power-law index changes, $V_\mathrm{break}$ is the rotational velocity at $R_\mathrm{break}$, and $p_\mathrm{in}$ and $p_\mathrm{out}$ are power-law indices at $r \leq R_\mathrm{break}$ and $r > R_\mathrm{break}$, respectively. Free parameters are $V_\mathrm{break}, R_\mathrm{break}, p_\mathrm{in}$, and $p_\mathrm{out}$. The parameter space of $\vsys$ is also explored from $7.18 \kmps$ to $7.26 \kmps$ with steps of $0.01 \kmps$, although it is not treated as a free parameter in each fitting to reduce the number of free parameters. The data points and fitting result are shown in Figure \ref{rplaw}b. The best fit parameters for the single power-law function are $(\vsys, V_{100}, p) = (7.20 \pm 0.02 \kmps, 3.74 \pm 0.20 \kmps$, and $0.60 \pm 0.03)$, giving reduced $\chi^2=8.3$. The best fit parameters for the double power-law function are $(V_\mathrm{break}, R_\mathrm{break}, p_\mathrm{in}, p_\mathrm{out}) = (1.31 \pm 0.08 \kmps, 624 \pm 62 \au, 0.47 \pm 0.05, 1.07 \pm 0.24)$ at $\vsys = 7.23 \kmps$, giving reduced $\chi^2 = 2.4$. Parameter errors are derived by the same method for the inner component, but Gaussian distributions are assumed for both position and velocity errors. The derived power-law indices of the rotational profile imply that the outer component is a Keplerian disk within $r\sim$600 au, which transitions to an infalling envelope, as observed in several other protostars \citep{Murillo:2013aa,Chou:2014aa,Ohashi:2014aa,Aso:2015aa,Aso:2017ab}. This kinematically derived disk radius is interestingly comparable to the outer edge of the \ce{C^18O} emission on the NE side (i.e., blueshifted side), which can be considered as the disk radius assuming axisymmetry in the disk. Note that the derived disk radius is much larger than the radii of the \ce{C^18O} dips.
%
% Kinetic Model
\subsection{Kinetic Model\label{kineticmodel}}
% ########### Table to present model parameters ##############
%
\begin{table}[thbp]
\begin{threeparttable}
    \begingroup
    \centering
    \caption{Parameters of the warped disk model}
    \begin{tabular*}{\columnwidth}{@{\extracolsep{\fill}}lc}\hline\hline
    Parameter & Value \\\hline 
    Central stellar mass ($M_\ast$) & $1.6 \Msun$ \\
    Disk mass ($M_\mathrm{disk}$) & $0.0071 \Msun$ \\
    Power of surface density distribution ($p$) & 0.5 \\
    Inclination angle ($i$) & $73\degree$\\
    \textbf{for inner disk} &  \\
    \ \ \ Disk inner radius ($R_\mathrm{in}$) & $ 0.1 \au$ \\
    \ \ \ Disk outer radius ($R_\mathrm{out}$) & $ 200 \au$ \\
    \ \ \ Position angle (P.A.) & 69\degree \\
    \textbf{for outer disk} & \\
    \ \ \ Disk inner radius ($R_\mathrm{in}$) & $ 300 \au$ \\ 
    \ \ \ Disk outer radius ($R_\mathrm{out}$) & $ 600 \au$ \\
    \ \ \ Position angle (P.A.) & 54\degree \\
    \hline
    \end{tabular*}
	% footnote of table
	\endgroup
\end{threeparttable}
\label{table_disk_param}
\end{table}
%
% ############ Figures: models ###########
% channel maps of the model
\begin{figure*}[thbp]
\centering
\begin{minipage}[ht]{1.55\columnwidth}
\centering
\includegraphics[width=1\columnwidth]{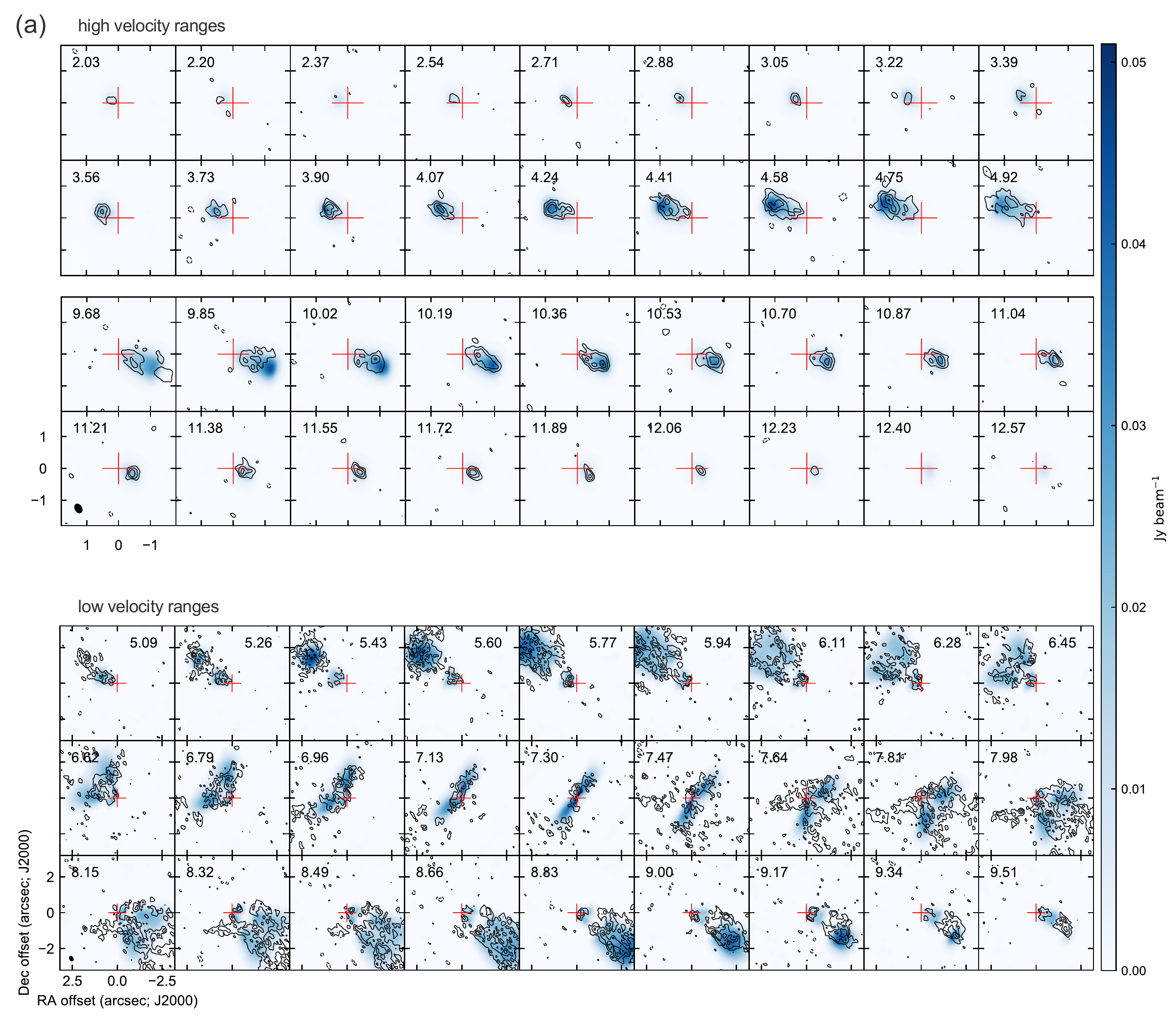} %  0 0 792 612
\end{minipage}\\ \vspace{9pt}
\begin{minipage}[ht]{1.55\columnwidth}
\centering
\includegraphics[width=1\columnwidth]{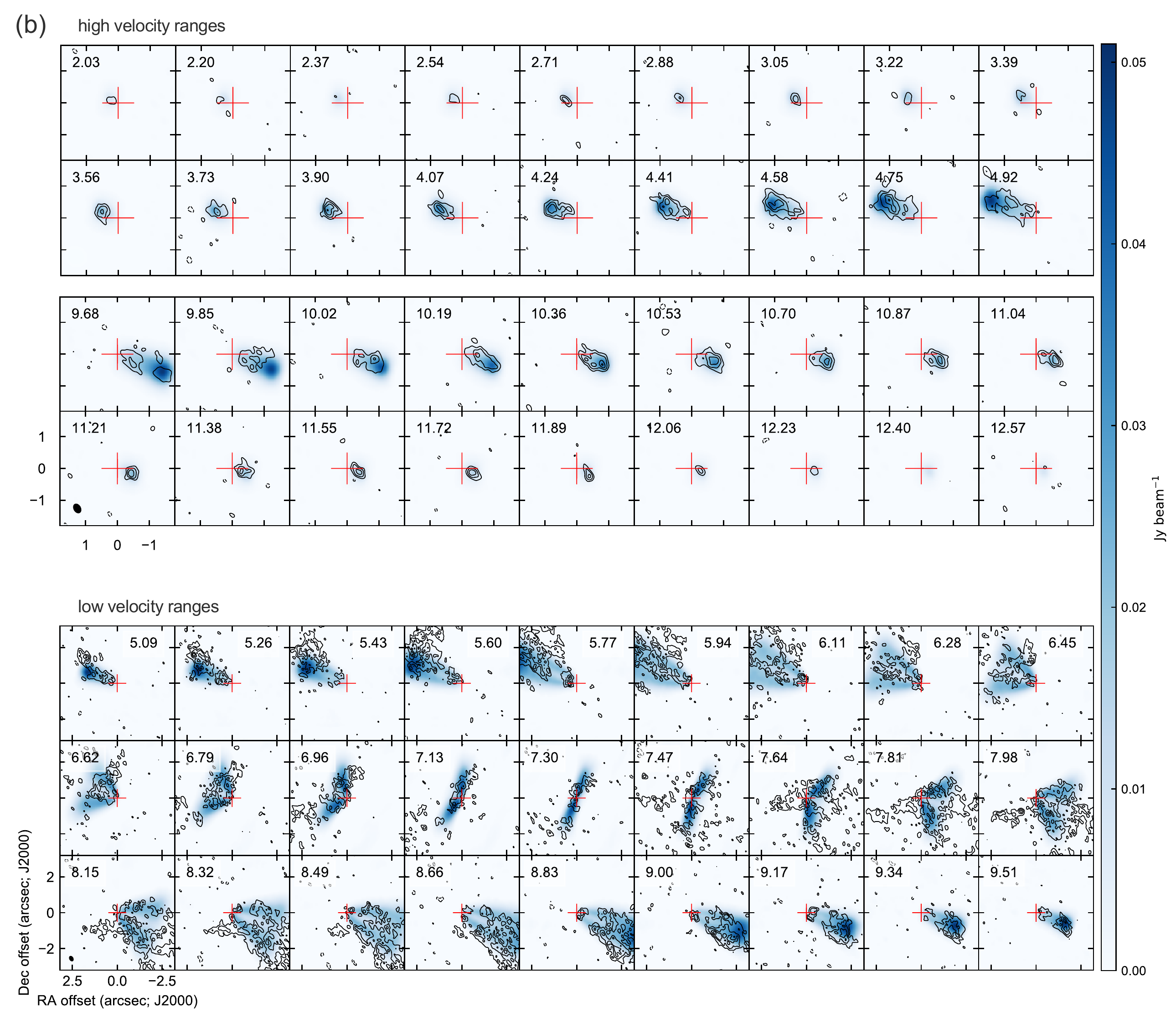} %  0 0 792 612
\end{minipage}
\caption{Same as Figure \ref{channelc18o21} but with the (a) warped disk model and (b) coplanar disk model in color (Section \ref{kineticmodel}). The \ce{C^18O} dips in 5.26--6.28$\kmps$ and 8.32--9.34$\kmps$ are reproduced by the warped disk model having a gap but not reproduced by the coplanar disk model. The coplanar disk model also shows small offset from the outer disk components of the observations in 5.26--6.45$\kmps$ and 8.15--9.17$\kmps$ in the P.A. direction.}
\label{channel-model}
\end{figure*}
% ###############################################
% contents
As shown in the previous sections, a Keplerian disk likely extends up to $r\sim 600$ au with \ce{C^18O} dips at radii of $\sim$280 au on the NE side and $\sim$140 au on the SW side of the protostar. Hereafter, we refer the inner and outer parts of the disk with respect to the \ce{C^18O} dips ($r\lesssim 200$ au and $r\sim$ 300--600 au) to as an inner disk and an outer disk, respectively. Notably, the position angles of the inner and outer disks are different by $\sim$15\degree, indicating that the rotational axis of the disk changes its direction with radii, i.e., the disk is warped. In this section, we model a warped disk and compare it with observations.

For the disk model, the following surface density profile is assumed:
\begin{align}
\Sigma (r) &= \Sigma_0 \left(
\frac{r}{1 \au}
\right)^{-p},
\end{align}
where $\Sigma_0$ satisfies the following equation:
\begin{align}
\int_{R_\mathrm{in}}^{R_\mathrm{out}} 2 \pi r \Sigma dr = M_\mathrm{disk}.
\end{align}
By assuming hydrostatic equilibrium, the vertical density distribution is expressed as
\begin{align}
\rho (r,z) = \frac{\Sigma (r)}{\sqrt{2\pi}h}
\exp \left( -\frac{z^2}{2h^2}
\right).
\end{align}
Here, $h$ is the scale height, which is defined as $h \equiv c_\mathrm{s}/\Omega$, where $ c_\mathrm{s}=\sqrt{kT/\mu m_\mathrm{H}}$ is the isothermal sound speed and $\Omega=\sqrt{GM_\ast/r^3}$ is the angular velocity. Keplerian rotation is assumed as the velocity field, and no vertical and radial motions are included. We use the 3D radiative transfer code RADMC-3D\footnote{\url{http://www.ita.uni-heidelberg.de/~dullemond/software/radmc-3d/}} \citep{Dullemond:2012aa} to calculate the temperature distribution and synthetic images of our model with $M_\ast = 1.6 \Msun$ and $T_\ast = 4000$ K. The scale height is given by $h/r = 0.03\ (r/ 1\au)^{0.25}$. We adopt the dust opacity calculated in \cite{Semenov:2003aa}\footnote{\url{https://www2.mpia-hd.mpg.de/home/henning/Dust_opacities/Opacities/opacities.html}}, using the composite aggregate dust model of the normal abundance at temperature of $T\lesssim 150$ K.

A warped disk model is produced by adding two separate disk models in the image domain: an inner disk and an outer disk. Although this method could be incorrect at optically thick regions, especially around the systemic velocity, this separate model construction is plausible because our models are basically optically thin. We also note that our model does not include the observed non-axisymmetric feature. No parameter fitting is performed because our purpose here is to confirm whether the warped disk model can reproduce the observed features rather than to constrain physical parameters of the disk. The warped disk model includes a gap structure, where material is absent, to reproduce the \ce{C^18O} dips. We set the radial ranges of the inner and outer disks as 0.1--200 au and 300--600 au, respectively, based on the \ce{C^18O} dip on the NE side from the observations. The inner and outer disks have different position angles of $69\degree$ and $54\degree$, respectively, as derived from Figure \ref{lowhighvelmom}. The inclination angles of the inner and outer disks are set to the same angle of $73\degree$. The parameters for the warped disk model are listed in Table \ref{table_disk_param}. For comparison, we also produced a coplanar disk model with no warped structure and no gap. The disk radius and position angle of the coplanar disk are 600 au and $69\degree$, respectively (i.e., aligned with the inner disk of the warped disk model). The temperature distribution of the coplanar disk calculated from RADMC-3D is used for both the inner and outer disks for the warped disk model. The produced disk models are observed using the CASA task \textit{simobserve} with the ALMA antenna configurations similar to those in our observations. The synthesized beam for the model observations is $\ang[angle-symbol-over-decimal]{;;0.35}\times \ang[angle-symbol-over-decimal]{;;0.24}\ (7.4\degree)$.

\begin{figure*}[thbp]
\centering
\includegraphics[width=1.5\columnwidth]{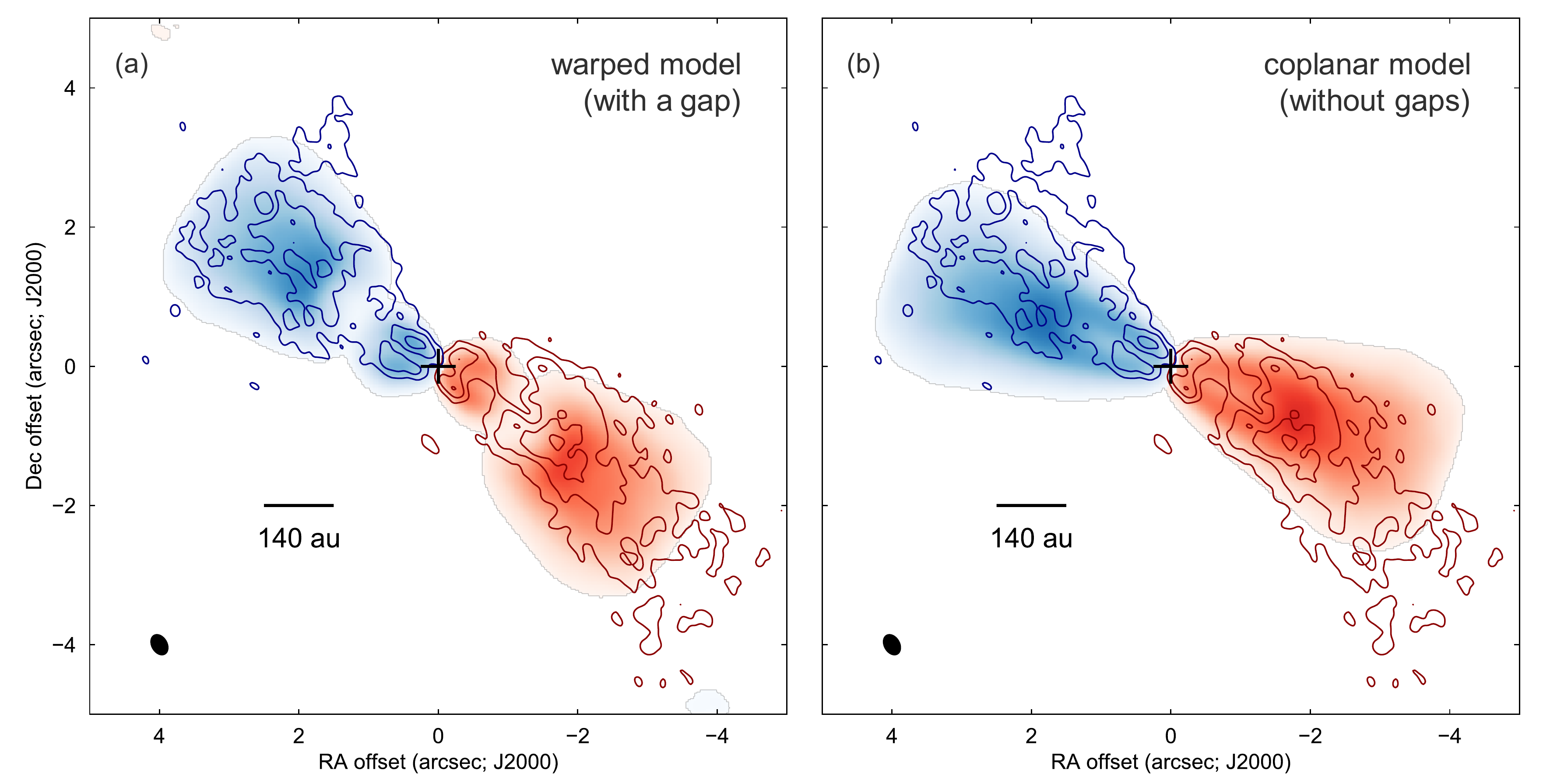} 
\caption{Comparison between the observed \ce{C^18O} $J=$2--1 moment 0 maps of the low-velocity components (\textit{contours}, same as those in Figure \ref{lowhighvelmom}b) to those of the warped and coplanar disk models (\textit{colors}). Color scales are shown from 0.7$\sigma$ to 12$\sigma$, where 1$\sigma$ corresponds to 3.6 $\mjpbm \kmps$ for the blueshifted component and 3.9 $\mjpbm \kmps$ for the redshifted component.}
\label{mom0_lowvel_model}
\end{figure*}

The velocity channel maps of the warped disk model are shown with observations in Figure \ref{channel-model}a. The warped disk model reproduces the high velocity components. The blueshifted component of the model at velocities of $\sim$5--7$\kmps$ matches that of the observations, including the \ce{C^18O} dips. The gap of the redshifted component of the model is located further than the \ce{C^18O} dips of the observations in the velocity range of $\sim$8.5--9.5$\kmps$, because the non-axisymmetric feature seen in the observations is not included in the model. Nevertheless, the warped disk model reproduces most of the observational features except for the non-axisymmetry. The velocity channel maps of the coplanar disk model are shown in Figure \ref{channel-model}b. The coplanar disk model matches the high velocity components as the warped disk model does. However, in the low velocity range of $\sim$5--9.5$\kmps$, the coplanar disk model does not reproduce the \ce{C^18O} dips. The model also shows a slight offset of the emission distribution in the azimuthal direction from that of the observations, especially in the channels of 5.43--6.28$\kmps$ and 8.32--8.66$\kmps$. Figure \ref{mom0_lowvel_model} compares the moment 0 maps of the low velocity components of the models and the observations. The warped disk model matches the observations, while the coplanar disk model exhibits a clear offset from the observations in the P.A. direction.
% ######### figure: PV diagram ###########
% % model moment 0 of low velocity comp
%
% model PV diagram
\begin{figure*}[thbp]
\centering
\includegraphics[width=1.5\columnwidth]{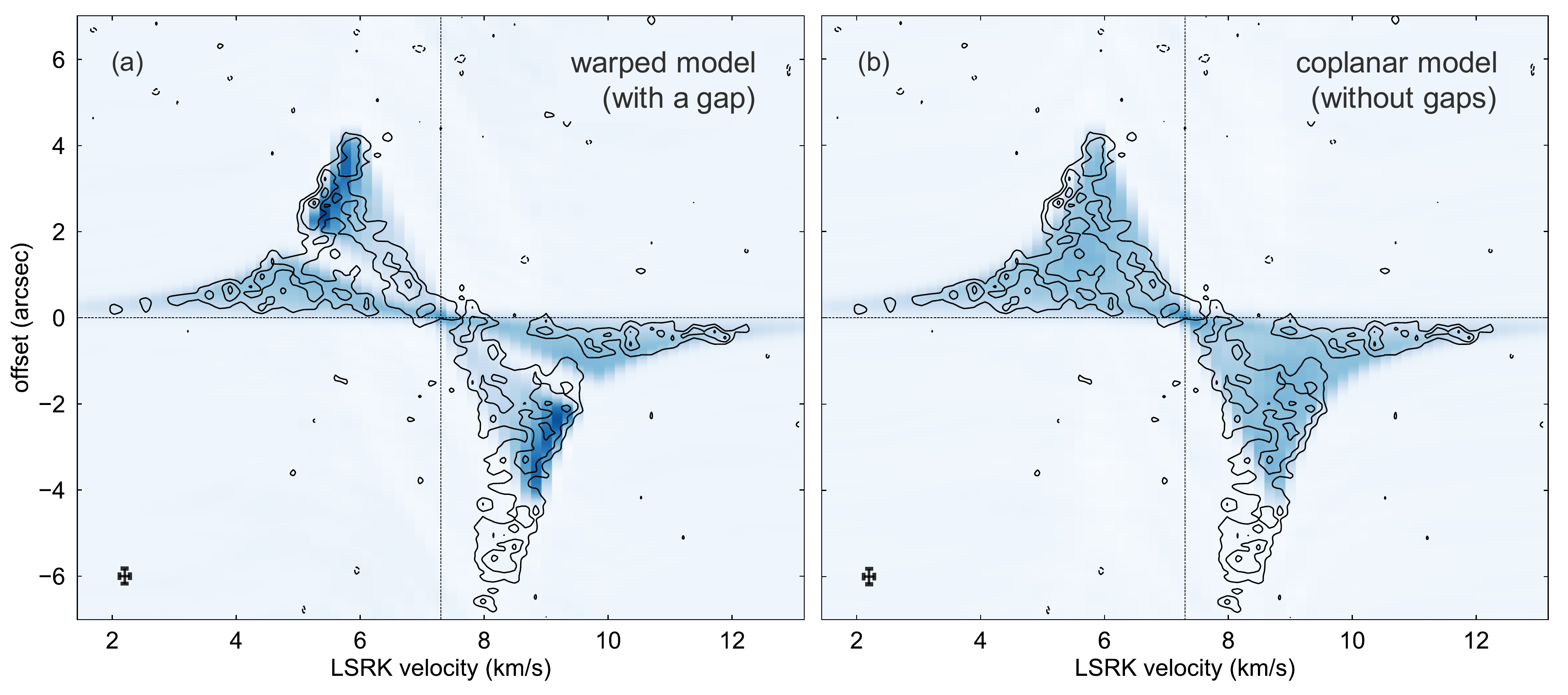} 
\caption{Comparison of the PV diagrams of the (a) warped disk model and (b) coplanar disk model (\textit{colors}) with the observed PV diagram (\textit{contours}) without tapering along P.A.$=54\degree$. Contour levels are from $3\sigma$ to $9\sigma$ in steps of $2\sigma$, where $1\sigma$ corresponds to 5.1 $\mjpbm$. Vertical and horizontal bars at the bottom-left corners denote the FWHM of the synthesized beam along its major axis and the velocity resolution.}
\label{pvmodel}
\end{figure*}
% ######################################
Figure \ref{pvmodel} shows the PV diagrams of the models and observations. Overall, the warped disk model shows the same features as those of the observations having Keplerian rotation and the \ce{C^18O} dips. The coplanar disk model does not produce the \ce{C^18O} dips in its PV diagram, which is also shown in the comparison with channel maps. These results suggest that the warped disk model explains the observational features better.

Note that both models have the outermost radius of 600 au even though the \ce{C^18O} emission extends beyond that radius. This means that both models cannot explain emission outside 600 an in radius. In fact, the emission at the offset of $\sim-\ang[angle-symbol-over-decimal]{;;6}$ and at the velocity of $\sim$8$\kmps$ in Figure \ref{pvmodel} is not explained by either model. Importantly, this part of the PV diagram corresponds to the part where rotation follows $v \propto r^{-1}$ as discussed in Section \ref{analysis_vc18o}. In another word, the emission in this part of the PV diagram should arise from an infalling envelope. An infalling envelope usually shows characteristic features in its PV diagram along its major axis, such as a large spread of emission in positions at near the systemic velocity \citep[e.g.,][]{Ohashi:2014aa,Aso:2015aa}. The PV diagram shown in Figure \ref{pvmodel} and also in Figure \ref{pvobs}, however, does not show such a feature. This could suggest that the infalling velocity is much smaller than the free-fall velocity and that the entire system of L1489 IRS is more rotation-dominant. In order to investigate this possibility, we have to perform careful measurements of infalling velocities in the system.
% ------------------------------------------------------------
%
% -------------- SECTION 5: Discussion ----------------
\section{Discussion\label{discussion}}
% Origin of the warp
\subsection{Morphology of the \ce{C^18O} emission}
% ##### Figures: Matsumoto-san's simulation ###
\begin{figure}[thbp] % xbb: 0 0 612 792
\centering
% surface density
\begin{minipage}[ht]{\columnwidth}
\centering
% viewport=0 150 612 692
\includegraphics[width=\columnwidth]{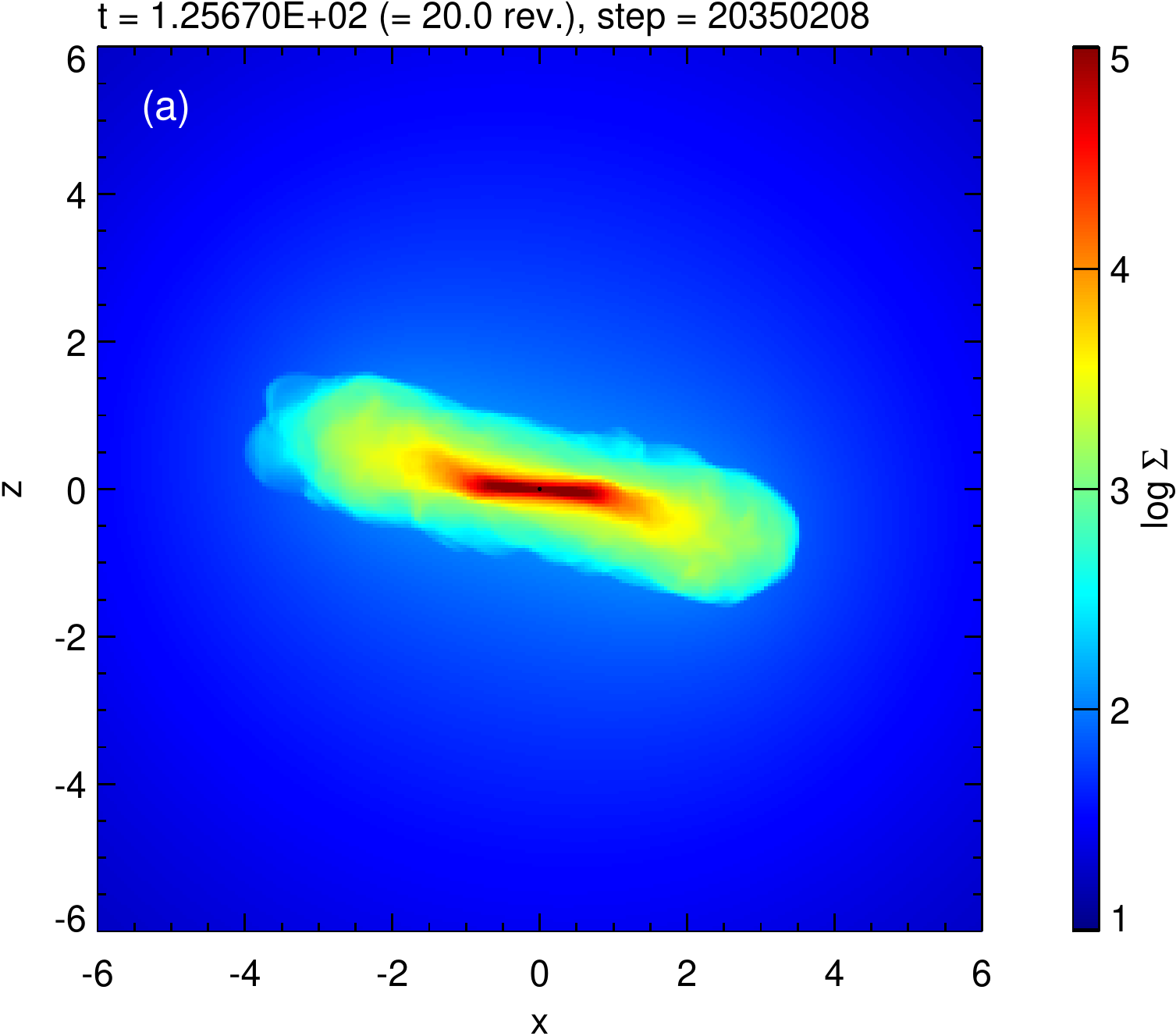}
\end{minipage}\\
% density
\begin{minipage}[ht]{\columnwidth}
\centering
\includegraphics[width=\columnwidth]{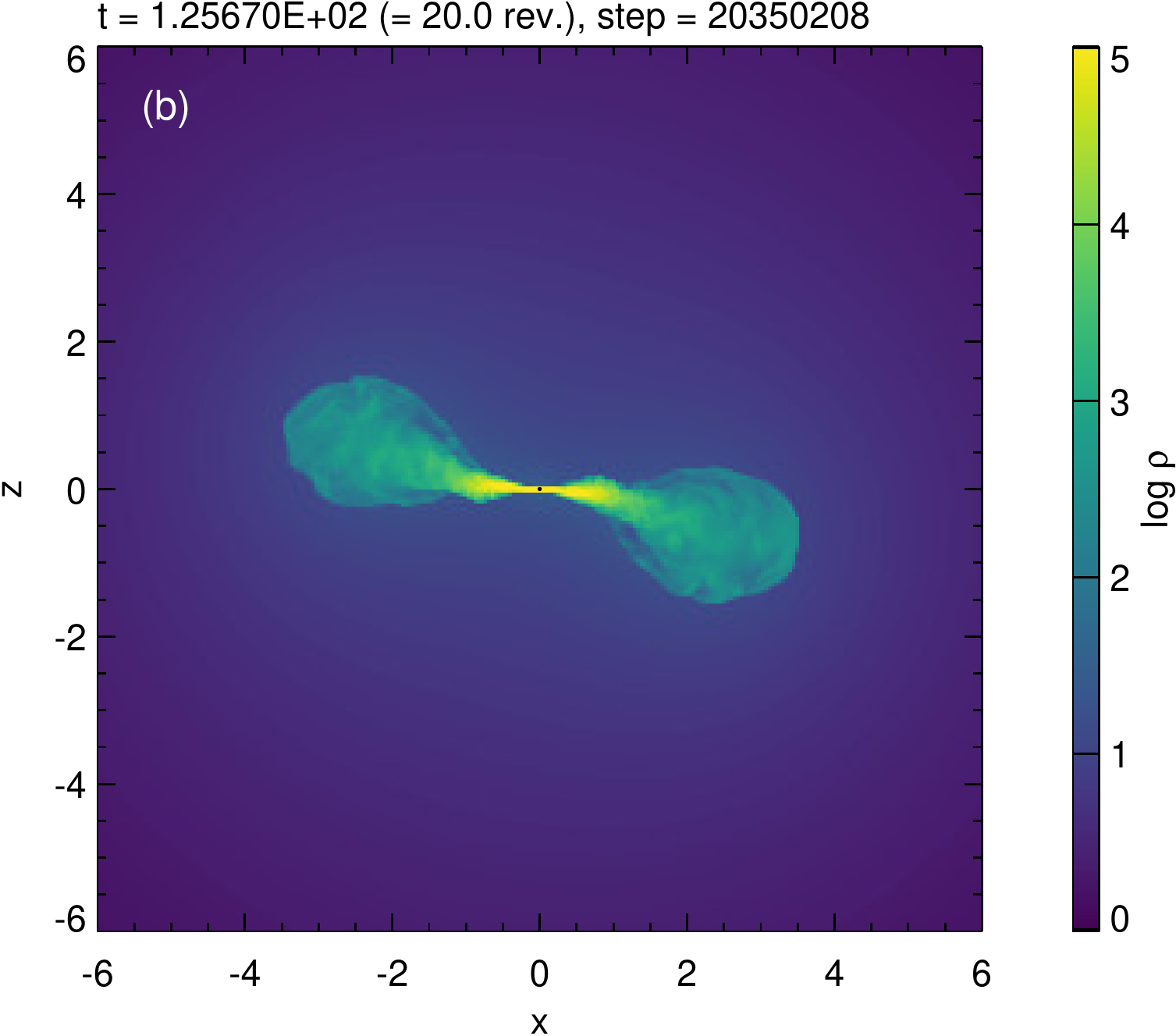}
\end{minipage}
\caption{Results of our numerical simulation of the warped disk model. (a) Surface density distribution along the y-axis and (b) density distributions in the $y=0$ plane at $t = 20 T_\mathrm{d}$. The spatial coordinates and density are normalized by $R_0$ and $\rho_0$, respectively (Appendix \ref{app_sim}). The small black circle in the center shows the sink radius.}
\label{result_sim}
\end{figure}
% ################################################
%
% ##### Relation to Yen+14 #####
\begin{figure}[thbp] % xbb: 0 0 612 792
\centering
% viewport=0 150 612 692
\includegraphics[width=\columnwidth,viewport=100 20 692 542]{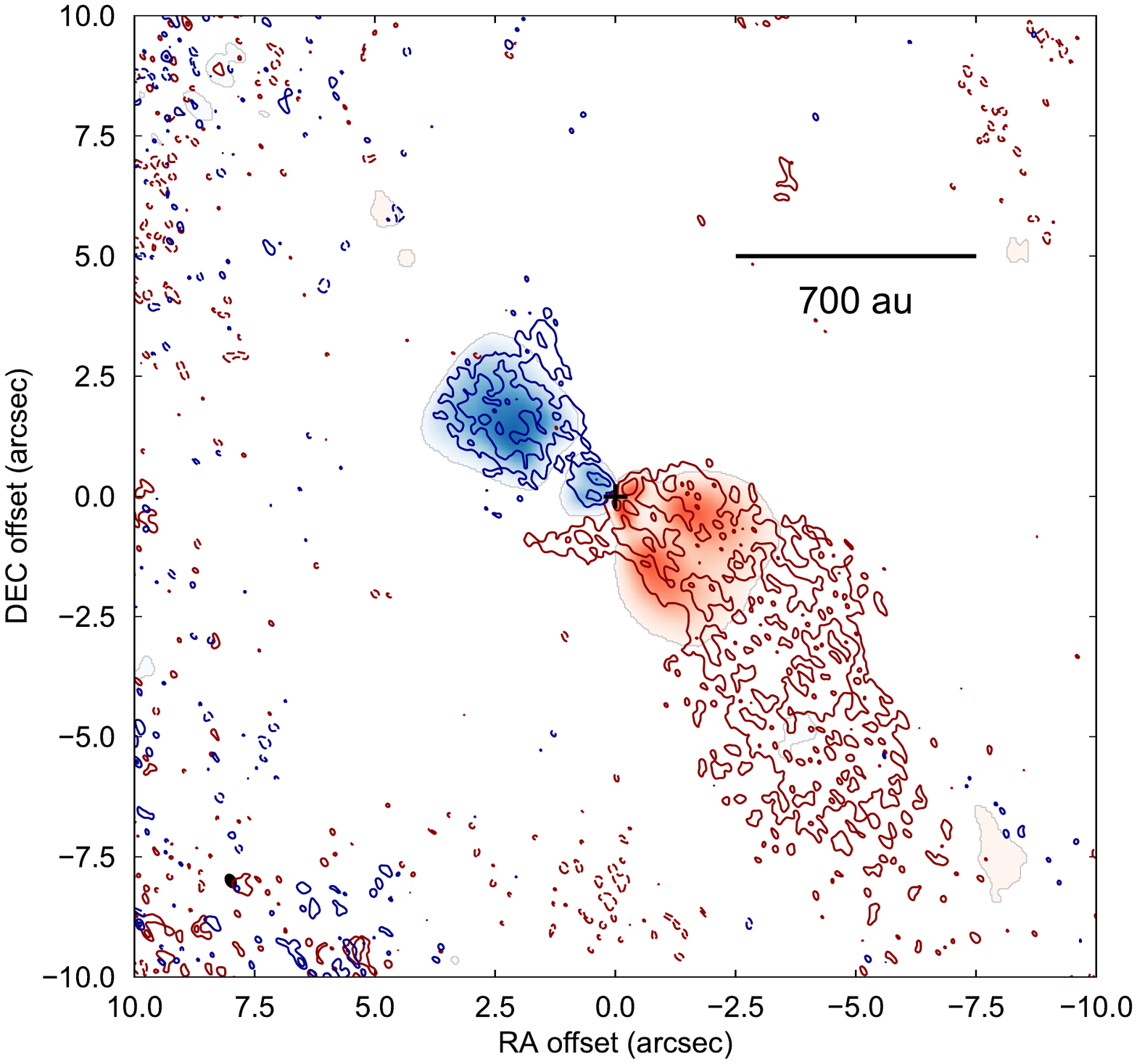}
\caption{Observed total intensity maps of the blueshifted and redshifted components integrated over the velocity ranges of $5.43\kmps$ to $6.00\kmps$ and $7.98 \kmps$ to $8.49 \kmps$, respectively (\textit{contours}) overlaid on the model counterparts (\textit{color}). The velocity ranges correspond to the low-velocity channels that show the most extended emission. Red and blue colors show the blueshifted and redshifted components, respectively. Contour levels are from $3\sigma$ to $7\sigma$ in steps of $2\sigma$, where $1\sigma$ corresponds to 2.7 $\mjpbm$.}
\label{fig_flows}
\end{figure}
% ############################################
% Relation between our model and Yen's infalling flows
% Shows emission which cannot be explained by our disk model
We have demonstrated that a warped disk can explain the observational features at $r \lesssim 600 \au$. In addition, outside the warped disk there is extended emission, which seems to arise from the infalling envelope. In this subsection, more details of the morphology of the \ce{C^18O} emission are discussed.

% subsub section: origin of warped disk
\subsubsection{Morphology of the Emission at $r<600$ au:\\ Warped Structure}
% Brinch's result
A disk and envelope structure similar to the warped disk has been suggested in L1489 IRS based on interferometric observations in molecular lines and continuum, and an SED \citep{Brinch:2007aa}. Their kinematic model suggests that a Keplerian disk with a radius of 200 au is misaligned with respect to an envelope by $\sim$30\degree. This is consistent with our results, but our analysis kinematically indicates that a part of a structure considered as an envelope in the previous work is a warped Keplerian disk.

% possibility that the inclination angles are also different between the inner and outer disks
Note that the inclination angle of the outer disk was set to the same as the inner disk in the warped disk model because it is difficult to measure the inclination angle of the outer disk independently of the inner disk from the current data set. However, not only the position angle but also the inclination angle could be different between the inner and our disks. The derived rotational velocity $V_\mathrm{break}$ in Section \ref{analysis_vc18o} allows us to estimate possible inclination angles of the outer disk as follows. While the rotational velocity at $R_\mathrm{break} \sim 620$ au is predicted as 1.54 $\kmps$ from the stellar mass $1.6 \Msun$, the derived $V_\mathrm{break}=1.31\ \kmps$ without the correction of the inclination angle is smaller than the prediction by $\sim$0.87. From these values and the relation $\vlsr = \vrot \sin i$, the inclination angle is estimated as $i\sim60\degree$. On the other hand, the difference between the predicted and derived rotational velocities can be interpreted as a consequence of a limited angular resolution and the integration effect along the line-of-sight in the observations as discussed in Appendix \ref{app_pv}. The same fitting method on a PV diagram is applied to a disk model with $i=90\degree$ and $M_\ast=1.6\ \Msun$, showing that the derived peak velocity is smaller than the rotational velocity by $\sim$0.87 (see Figure \ref{rn_vs_vn_compdv}). This is because the spectra are skewed due to the beam smearing and the integration along the line-of-sight, and thus the derived peak velocity shifts to the lower-velocity side. This examination suggests that $i=90\degree$ also explains the derived rotational velocity. Therefore, the outer disk can have an inclination angle different from that of the inner disk within a range of $\sim$60\degree--90\degree.

% Turbulence Scenario: Matsumoto-san simulation
One possible cause of such a warped disk structure is that the infalling envelope in L1489 IRS has changed the angular momentum axis during its accretion process. Numerical calculations simulating star and disk formation in a core with an initial condition, where the core is magnetized and has a random velocity field due to turbulence, suggest that a rotational axis of the envelope is aligned with the magnetic field vector, while that of the inner disk is along the angular momentum vector, resulting in misalignment between the disk and envelope \citep{Matsumoto:2017aa}. This result suggests that a warped outer disk can form by accretion from an envelope having an angular momentum axis different from that of an inner disk. To confirm that such an accretion results in a warped disk system, we performed a hydrodynamics simulation calculating disk formation from infalling gas whose angular momentum varies its direction at a certain time in the simulation (see Appendix \ref{app_sim}). Figure \ref{result_sim} shows the results of the simulation, demonstrating that a warped disk can form. Accretion forms an inner disk parallel to the $x$ axis at an early phase, followed by a warped outer disk tilted from the $x$ axis after changing the direction of the angular momentum axis of the envelope.

% Bate
Other numerical simulation works have shown similar results. A simulation of disk formation in a turbulent cloud demonstrates that misaligned disks form around binaries through accretions changing their angular momentum axes during the disk formation process because of the turbulent environment \citep{Bate:2018aa}.
% Magnetic field effect: Hiroano-san results
Misalignment between an angular momentum axis of a core and a magnetic field vector can also cause the warped structure in a protostellar system. Simulation works calculating star formation in such a core showed that an outer lower-density region, corresponding a pseudo-disk, is along with the magnetic field at the early phase, whereas at the later phase the magnetic field dissipates in an inner high-density region and a disk forms around the angular momentum axis \citep{Matsumoto:2004aa,Hirano:2019aa}.

% torque effect: Facchini
In a binary system where a binary has an orbit misaligned from the equatorial plane of its circumbinary disk, binary torque can tear the disk, resulting in misalignment between an inner disk and outer disk, or more generally, a warped disk \citep{Nixon:2013aa}. \cite{Nixon:2013aa} calculated such a situation theoretically and derived formulae predicting the disk breaking radius $r_\mathrm{break}$. \cite{Facchini:2018aa} performed a numerical simulation calculating the evolution of a circumbinary disk misaligned from the binary orbit; they confirmed that a circumbinary disk is broken into two misaligned disks, and the break radius is consistent with the theoretical prediction in \cite{Nixon:2013aa}. It has been speculated that L1489 IRS is a binary system \citep{Hogerheijde:2000aa, Brinch:2007aa}, although there is still no direct evidence for its binarity. A scattered light image of L1489 IRS shows that it is a point source at an angular resolution of $\sim \ang[angle-symbol-over-decimal]{;;0.2}$ ($\sim$30 au) \citep{Padgett:1999aa}. Thus, the separation of the binary would be less than 30 au even if L1489 IRS is a binary.

% estimate rbreak due to the binary torque
Assuming a couple of parameters of a binary and a disk, an upper limit of $r_\mathrm{break}$ due to the binary torque can be estimated based on theoretically derived equations in two different regimes; the bending wave regime ($h/r \gg \alpha$) where pressure provides the dominant internal disk torque, or the diffusive regime ($h/r < \alpha < 1$) where viscosity provides the dominant internal disk torque \citep{Nixon:2013aa,Nixon:2016aa}. In the case of L1489 IRS, the warped structure appears to break at the \ce{C^18O} dips at $r \sim$ 200--300 au. Its disk radius of $\sim$600 au and the breaking radius suggest that the disk angular thickness $h/r$ would be large if the disk satisfied the hydrostatic equilibrium, and thus it would be in the bending wave regime. Indeed, assuming a typical temperature profile of $T(r) = 400 (r/\au)^{-0.5}$, $h/r$ is estimated to be $\sim$0.1 at 100 au, which is larger by one order of magnitude than the typical value of $\alpha$ viscosity, $\alpha = 10^{-2}$, in a protoplanetary disk. Given $h/r \gg \alpha$, the limit of $r_\mathrm{break}$ is estimated as $r_\mathrm{break} \lesssim 2 a = 60 \au$ from Equation (A3) in \cite{Nixon:2013aa}, assuming $\mu=M_\mathrm{s}/(M_\mathrm{p}+M_\mathrm{s})=0.5$, where $M_\mathrm{p}$ and $M_\mathrm{s}$ are the masses of the primary and secondary stars, respectively, an inclination angle between the binary and disk of $\theta = 45\degree$, separation of the binary of $a=30 \au$, and $h/r=0.1$. Note that the assumed $\mu$ and $\theta$ maximize the equation and thus give the upper limit of $r_\mathrm{break}$, although these two parameters of the binary are uncertain. This upper limit implies that the binary torque is unlikely to break the disk at a large radius of $\sim$200--300 au, even if L1489 IRS is a binary.
%
% subsub section: relation to Yen+14's work
\subsubsection{Morphology of the Emission at $r>600$ au}
Figure \ref{fig_flows} shows the distribution of the most extended emission over the velocity rages of $5.43\kmps$ to $6.00 \kmps$ and $7.98 \kmps$ to $8.49 \kmps$, which is not explained by the warped disk model. The redshifted component extending up to $r\sim 1200$ au has also been observed in the previous observations at a lower resolution of $\sim\ang[angle-symbol-over-decimal]{;;1}$ in the \ce{C^18O} $J=$2--1 line \citep[see Figure 13 in][]{Yen:2014aa}, demonstrating that it exhibits arm-like shape although such a shape is not very obvious in our image. A similar but small arm-like structure has been seen as a blueshifted component on the northeastern side of the disk component in the previous observations while it is not detected in the current observations at a higher angular resolution due to the lack of SNR.

% review of Yen+14
The extended, arm-like structures were interpreted as two stream-like infalling flows accreting at $\sim$300 au in radius on a disk with the outermost radius of $\sim$700 au \citep{Yen:2014aa}. The mass of the infalling flows is estimated to be 4--7$ \times 10^{-3} \Msun$ from the total integrated \ce{C^18O} flux assuming the LTE condition. Based on the mass and the infalling timescale derived from the lengths and infalling velocities of the flows, the infalling rate is estimated to be $\sim$4--7$ \times 10^{-7} \msunpyr$. They found that the estimated infall rate is larger than the rate of mass accretion onto the central star by a factor of 2--4, suggesting that the disk of L1489 IRS may continue to obtain mass from its envelope. The infalling flows also imply that the disk size can expand via redistribution of the angular momentum within the disk because the infalling flows accrete on the disk plane at $r\sim 300 \au$. The landing radius is, however, not well-constrained due to the limited angular resolution and assumed based in the intensity peaks of SO and \ce{C^18O} emissions. As well as the landing radius, the direction of the angular momentum axis of the infalling flows is also not well determined while it is assumed to be the same as that of the disk. The infalling flows could be an origin of the warped disk structure if it is landing at the edge of the disk and has an angular momentum axis in the different direction from that of the disk.
%
% Origin of the CO dip
\subsection{\ce{C^18O} Dips\label{origin_of_gap}}
While the \ce{C^18O} emission shows dips, as described in Sections \ref{results} and \ref{analysis}, no such counterpart appears in the 1.3 mm dust continuum emission. Similar dips are also observed in the \ce{C^17O} $J=$2--1 line emission (van't Hoff et al. 2020 in prep). The observed \ce{C^18O} dips are thus likely due to the deficient CO abundance rather than the physical absence of material.

% freeze -out
\ce{CO} molecules can deplete by freeze-out at temperatures below $\sim$20--30 K in the typical disk density range \citep{Furuya:2014aa}. Producing the \ce{C^18O} dips by CO freeze-out requires a non-monotonic temperature distribution to decrease the disk temperature only at the dip radii. A few works suggest that a non-uniform size distribution of dust grains in a disk, which is caused by interactions between dust grains and gas (e.g., radial drift), can produce a non-monotonic disk temperature profile \citep{Cleeves:2016aa,Facchini:2017aa}. In such a situation, CO molecules freeze-out at a certain radius in a disk, while they can be thermally desorbed again at larger radii due to a radial thermal inversion in a disk.

% non-thermal desorption
\ce{CO} molecules could also be non-thermally desorbed at radii larger than the CO snow line even if a temperature profile is monotonic. A double ring of the \ce{DCO+} emission, suggesting a double \ce{CO} snowline, has been observed in a disk around the T Tauri star IM Lup, and it can be explained by non-thermal desorption of CO molecules at an outer region of the disk \citep{Oberg:2015aa}.

% temperature of the disk of L1489 IRS
The temperature profile of the disk and the envelope around L1489 IRS, which explains the spectral energy distribution of L1489 IRS, suggests that the temperature at 200 au is $\sim$30 K \citep{Brinch:2007aa}. The disk temperature at 200 au is calculated as $\sim$20--30 K with RADMC-3D in our disk model, which does not include an envelope. These results imply that the disk temperature could be $\sim$20--30 K around the \ce{C^18O} dip radius and thus \ce{CO} molecules could freeze-out. Detailed investigations of the temperature profile of L1489 IRS are needed to confirm whether the \ce{CO} freeze-out scenario can be the origin of the observed \ce{C^18O} dips.
%
% Comparison between Class I and Class II sources
\subsection{Comparison with Disks around other Protostars and T Tauri Stars}
%
% ##### Tables: samples ##########
% ### Table for Class I Sourecs ###
\begin{table*}
\begin{threeparttable}
    \begingroup
    \centering
    \scalefont{0.85}
    \caption{Parameters of Class I Protostars}
    \begin{tabular*}{2\columnwidth}{@{\extracolsep{\fill}}lccccccccc}\hline\hline
    Source & $M_\ast\ (M_\odot)$ & $ R_\mathrm{disk}\ (\mathrm{au})$ & $M_\mathrm{disk}\tnote{a}\  (M_\odot)$ & $L_\mathrm{bol}\ (L_\odot)$ & $T_\mathrm{bol}\ (\mathrm{K})$ & $L_\mathrm{bol}/L_\mathrm{submm}$ & $\dot{M}_\mathrm{acc}\ (M_\odot\ \mathrm{yr}^{-1})$ & $\dot{M}_\mathrm{inf}\ (M_\odot\ \mathrm{yr}^{-1})$ & References\\ \hline
    HH111 & $1.8 \pm 0.5$ & 160 & 0.14 & 23 & 78 & 100 & $1.2 \times 10^{-6}$  & $4.2 \times 10^{-6}$  & (1), (2), (3), (12) \\
    TMC-1A & $ 0.68 \pm 0.05$ & 100 & 0.0025 & 2.5 & 164 & $143$ & $3.5 \times 10^{-7}$ & 1.5--3.0$\times 10^{-6}$ & (4), (13) \\
    L1551 IRS 5 & $0.5^{+0.6}_{-0.2}$ & 64 & 0.070 & 24.5 & 106 & 107 & $4.7 \times 10^{-6} $ & $6 \times 10^{-6}$  & (5), (6), (13) \\
    IRS 43 & 1.9 & 700 & 0.0040 & 6.0 & 310 & - & $ 3.0 \times 10^{-7}$ & - & (7), (14), (15) \\
    L1489 IRS & $1.64 \pm 0.12$ & 600 & 0.0071 & 3.5 & 226 & 259 & $2.4 \times 10^{-7}$ & 4--7$\times 10^{-7}$  & this work, (8), (13) \\
    L1551 NE & $ 0.8^{+0.6}_{-0.4}$ & 300 & 0.047 & 4.2 & 91 & 100 & $5.0 \times 10^{-7}$ & $9.6 \times 10^{-7}$  & (9), (10), (12) \\
    IRS 63 & 0.8 & 165 & 0.099 & 1.5 & 287 & 33 & $ 1.8 \times 10^{-7}$ & - & (7), (13) \\
    TMC 1 & $0.54^{+0.2}_{-0.1}$ & 100 & 0.0039 & 0.66 & 171 & 114 & $ 1.2 \times 10^{-7} $ & - & (11), (13), (16) \\
    TMR 1 & 0.7\tnote{b} & 50 & 0.011 & 2.6 & 140 & 734 & $ 3.6 \times 10^{-7}$ & - & (11), (13), (16) \\
    L1536 & 0.4 & 80 & 0.021 & 0.6 & 270 & 125 & $ 1.2 \times 10^{-7}$ & - & (11), (17)\\ \hline
    \end{tabular*}
	% footnote of table
	\begin{tablenotes}
	\item \textbf{Notes.}
	\item[a] Listed values are taken from references. The values other than those for HH111, TMC-1A, L1489 IRS and L1551 NE are multiplied by 0.5 to account for the differences in dust opacity used to derive the disk mass in each paper.
	\item[b] The mass was derived from the bolometric luminosity.
	\item \textbf{References: }(1) \cite{Lee:2010aa};  (2)\cite{Lee:2011aa}; (3) \cite{Lee:2016aa}; (4) \cite{Aso:2015aa}; (5) \cite{Momose:1998aa}; (6) \cite{Chou:2014aa}; (7) \cite{Brinch:2013aa}; (8) \cite{Yen:2014aa}; (9) \cite{Takakuwa:2012aa}; (10) \cite{Takakuwa:2013aa}; (11) \cite{Harsono:2014aa}; (12) \cite{Froebrich:2005aa}; (13) \cite{Green:2013aa}; (14) \cite{Jorgensen:2009aa}; (15) \cite{Evans:2009aa}; (16) \cite{Kristensen:2012aa}; (17) \cite{Young:2003aa}.
	\end{tablenotes}
	\endgroup
\end{threeparttable}
\label{table_ClassIs}
\end{table*}
%
% Table for T Tauri Stars
\begin{table*}
\begin{threeparttable}
    \begingroup
    \centering
    \caption{Parameters of Class II Sources.}
    \begin{tabular*}{2\columnwidth}{@{\extracolsep{\fill}} lcccccc}\hline\hline
    Source & $M_\ast\ (M_\odot)$ & $R_\mathrm{disk}\ (\mathrm{au})$ & Binary separations\tnote{a}\ (arcsec) & References\tnote{b} & $ F_\mathrm{1.3mm}\tnote{c} \ (\mathrm{Jy})$ & $ L_\ast\tnote{c}\ (L_\odot)$ \\\hline
    DM Tau & $ 0.60 \pm 0.09$ & $641 \pm 19$ & s & G14 & $ 0.0894 \pm 0.0031$ & $0.226 \pm 0.016 $ \\
    MWC 480 & $ 1.83 \pm 0.37$ & $ 539 \pm 39$ & s & G14 & $ 0.2566 \pm 0.0093 $ & $19 \pm 12$ \\
    LkCa 15 & $ 1.09 \pm 0.07$ & $ 567 \pm 39$ & s & G14 & $ 0.1270 \pm 0.0049 $ & $ 0.81 \pm 0.36$ \\
    CI Tau & $ 0.80 \pm 0.02$ & $ 520 \pm 13$ & s & G14 & $ 0.1057 \pm 0.0057 $ & $ 0.93 \pm 0.30$ \\
    GO Tau & $ 0.48 \pm 0.01$ & $587 \pm 55$ & s & G14 & $ 0.0532 \pm 0.0028 $ & $ 0.285 \pm 0.080$ \\
    HV Tau C & $ 1.59 \pm 0.08$ & $256 \pm 51$ & s & G14 & $ 0.0231 \pm 0.0026 $ & $ 0.0274 \pm 0.0075$ \\
    DL Tau & $ 0.91 \pm 0.02 $ & $463 \pm 6 $ & s & G14 & $ 0.1688 \pm 0.0108$ & $ 0.74 \pm 0.31$ \\
    IQ Tau & $ 0.79 \pm 0.02 $ & $ 225 \pm 21$ & s & G14 & $ 0.0619 \pm 0.0045 $ & $ 0.81 \pm 0.22$ \\
    DN Tau & $ 0.95 \pm 0.16$ & $ 241 \pm 7 $ & s & G14 & $ 0.0823 \pm 0.0045 $ & $ 0.79 \pm 0.15$ \\
    FX Tau A & $ 1.70 \pm 0.18 $ & $ 40 \pm 10 $ & 0.68 & S17 & $ 0.0092 \pm 0.0043$ & $ 0.52 \pm 0.27$ \\
    CIDA-9 A & $ 1.08 \pm 0.2$ & $ 192 \pm 2 $ & 2.2 & S17 & $ 0.0295 \pm 0.0133 $ & $ 0.098 \pm 0.057$ \\
    HK Tau A & $ 0.58 \pm 0.05$ & $ 90 \pm 10 $ & 2.3 & S17 & $ 0.0313 \pm 0.0022 $ & $ 0.44 \pm 0.13$ \\
    HK Tau B & $ 1.00 \pm 0.03$ & $ 120 \pm 10 $ & 2.3  & S17 & $ 0.0145 \pm 0.0018 $ & $ 0.027 \pm 0.012$ \\
    IT Tau B & $ 0.50 \pm 0.08$ & $ 50 \pm 3 $ & 2.4 & S17 & $ 0.0042 \pm 0.0022 $ & $ 0.207 \pm 0.079$ \\
    DK Tau A & $ 0.60 \pm 0.14 $ & $ 38 \pm 4$ & 3.4 & S17 & $ 0.0166 \pm 0.0033 $ & $ 1.32 \pm 0.57$ \\
    GK Tau & $ 0.79 \pm 0.07$ & $82 \pm 2 $ & s & S17 & $ 0.0030 \pm 0.0010 $ & $ 1.35 \pm 0.73$ \\
    HN Tau A & $ 1.57 \pm 0.15$ & $ 52 \pm 10 $ & 3.0 & S17 & $ 0.0130 \pm 0.0060 $ & $ 0.42 \pm 0.35$ \\
    V710 Tau A & $ 0.66 \pm 0.06$ & $ 82 \pm 6 $ & 3.2 & S17 & $ 0.0299 \pm 0.0045 $ & $ 0.57 \pm 0.16$ \\
    HO Tau & $ 0.37 \pm 0.03$ & $ 62 \pm 5 $ & s & S17 & $ 0.0177 \pm 0.0033 $ & $ 0.130 \pm 0.026$ \\
    DS Tau & $ 0.73 \pm 0.02$ & $ 164 \pm 2 $ & s & S17 & $ 0.0165 \pm 0.0018 $ & $ 0.76 \pm 0.28$ \\
    CY Tau & $ 0.31 \pm 0.02$ & $ 290 \pm 10 $ & s & S17 & $ 0.0794 \pm 0.0059 $ & $ 0.400 \pm 0.083$ \\
    CX Tau & $ 0.37 \pm 0.02$ & $ 160 \pm 20 $ & s & S17 & $ 0.0094 \pm 0.0023 $ & $ 0.371 \pm 0.056$ \\
    IP Tau & $ 0.95 \pm 0.05$ & $ 95 \pm 20 $ & s & S17 & $ 0.0088 \pm 0.0015 $ & $ 0.47 \pm 0.14$ \\
    FM Tau & $ 0.36 \pm 0.01$ & $ 50 \pm 2 $ & s & S17 & $ 0.0131 \pm 0.0027 $ & $ 0.40 \pm 0.11$ \\\hline
    \end{tabular*}
	% footnote of table
	\begin{tablenotes}
	\item[a] "s" represents a single star.
	\item[b] G14: \cite{Guilloteau:2014aa}; S17: \cite{Simon:2017aa}.
	\item[c] From \cite{Andrews:2013aa}.
	\end{tablenotes}
	\endgroup
\end{threeparttable}
\label{table_ClassIIs}
\end{table*}
% ##### Figures: histogram ###
% disk size & mass distributions
\begin{figure*}[thbp] % xbb: 0 0 842 595
\centering
\includegraphics[width=1.8\columnwidth]{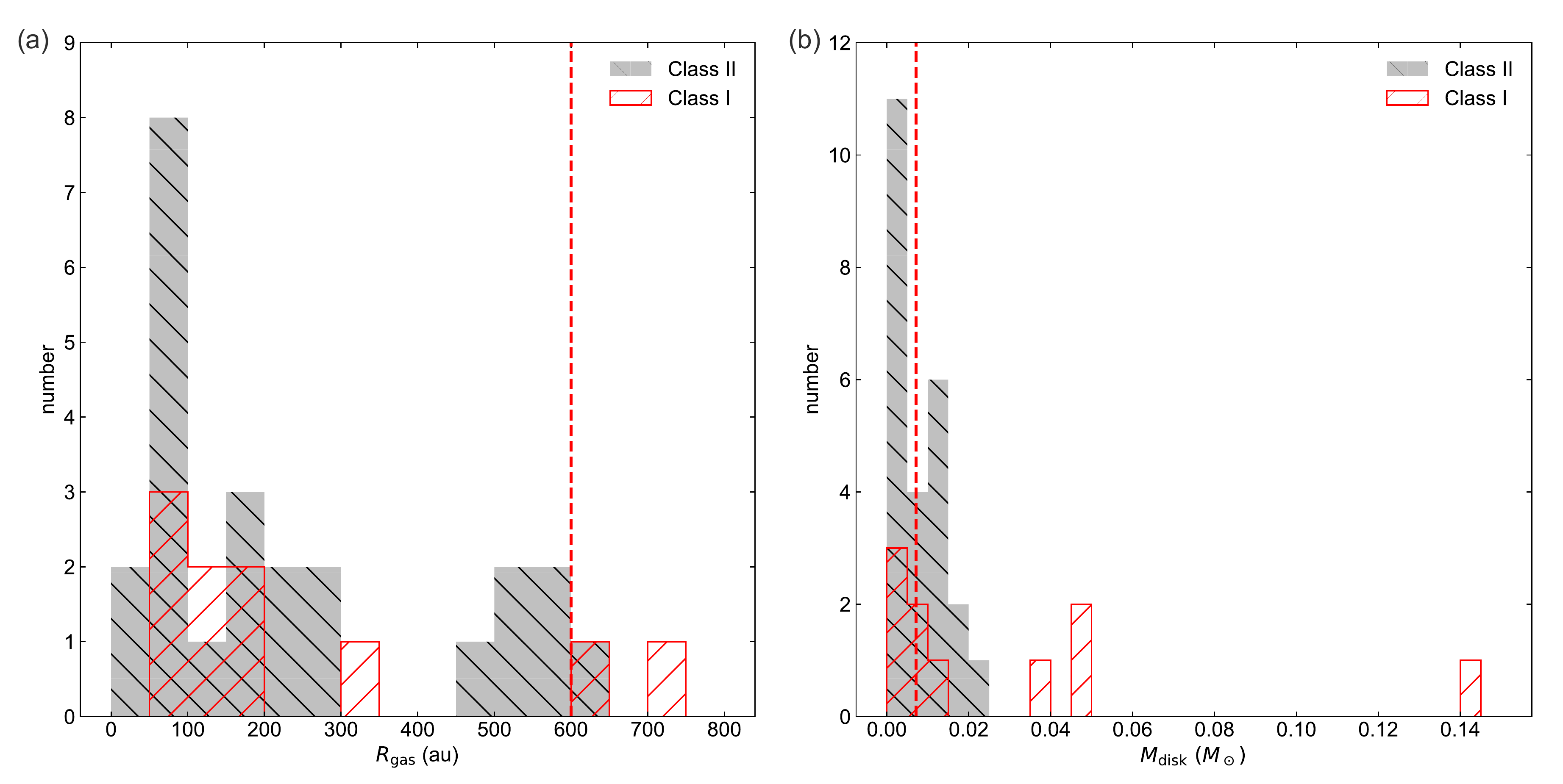}
\caption{Histograms of (\textit{left}) gas disk radii and (\textit{right}) disk masses of the Class I and Class II sources. The data values are listed in Table \ref{table_ClassIs} and \ref{table_ClassIIs}. Red vertical dashed lines denote the gas disk radius and disk mass of L1489 IRS. Note that disk masses in the table are multiplied by 0.5 and shown in the figure except for HH111, TMC-1A, L1489 IRS and L1551 NE in order to account for the different dust mass opacities used to derive disk mass in each paper.}
\label{disk_histo}
\end{figure*}
% #############################
\subsubsection{Disk Radius and Mass}
The disk radius of L1489 IRS ($\sim$600 au) is larger than that of most Class 0 or I protostars measured so far, although only few have been measured \citep{Yen:2017aa,Sheehan:2017aa}. In addition, the disk mass of L1489 IRS is small ($\sim$0.0071 $\Msun$) despite its large disk radius.

% sample description
We compare the disk radius and disk mass between Class I and Class \II sources to explore whether the physical disk parameters of L1489 IRS are attributable to evolutionary stages. The stellar and disk parameters of all sources discussed here are listed in Table \ref{table_ClassIs} and \ref{table_ClassIIs}. The Class I samples are taken from individual studies, which determine the central stellar mass kinematically except for TMR 1, whose stellar mass is estimated from its bolometric luminosity \citep{Harsono:2014aa}. The disk radius here means a gas disk radius and not a dust disk radius because the dust continuum lacks kinematic information, making it difficult to measure the dust disk radius accurately. The Class \II samples are taken from \cite{Guilloteau:2014aa,Simon:2017aa}, which measure gas disk radii. We compare only low-mass sources satisfying $M_\ast < 2\Msun$. The disk masses of the Class \II sources are derived from Equation (\ref{equation_diskmass}) using the flux densities measured at the 1.3 mm continuum in \cite{Andrews:2013aa}. We assume dust opacity of $\kappa_\mathrm{\nu} = 2.3\ \mathrm{cm^2 g^{-1}}$ at 1.3 mm \citep{Beckwith:1990aa} and $\langle T_\mathrm{disk} \rangle \sim 25(L_\ast/L_\odot)^{1/4}\ \mathrm{K}$ following \cite{Andrews:2013aa}. The disk masses around the Class I sources were derived using opacity $\kappa_\mathrm{\nu} = 1.8\ \mathrm{cm^2 g^{-1}}$ at 890 $\micron$ and $0.90\ \mathrm{cm^2 g^{-1}}$ at 1.3 mm suggested in \cite{Ossenkopf:1994aa}, except for HH 111, TMC-1A, L1489 IRS, and L1551NE, where the adopted dust mass opacities are smaller and the estimated disk masses are lager by a factor of 2--2.5 in \cite{Beckwith:1990aa}. To account for the difference in opacity, the disk masses of the Class I sources are multiplied by 0.5, except for those of HH 111, TMC-1A, L1489 IRS, and L1551NE, and plotted in the figures.

% stellar mass, disk raidus and disk mass correlation
% about the disk mass and radius of L1489 IRS
Figure \ref{disk_histo}a shows distributions of the gas disk radii ($R_\mathrm{gas}$) of the Class I and Class \II sources. The ranges and distributions of $R_\mathrm{gas}$ are similar between Class I and \II sources. \cite{Najita:2018aa} showed a result suggesting that the disk radius becomes larger from the Class I to Class \II stages due to viscous spreading, which contradicts our results. However, their Class I source sample includes a few Class 0 sources, which could still be increasing their disk radii by infall from envelopes \citep{Yen:2017aa}. The disk around L1489 IRS is within ranges of the distributions, showing that the disk radius is not especially large, although it is larger than most disks around other Class I sources.

Figure \ref{disk_histo}b shows distributions of the disk masses ($M_\mathrm{disk}$). The Class I protostars tend to have more massive disks than Class II sources, which was also reported in \cite{Sheehan:2017aa, Tychoniec:2018ab}. The disk mass of L1489 IRS is close to the mean value of Class \II sources. Therefore, the small disk mass around L1489 IRS would be due to the transitional phase of L1489 IRS from Class I to Class \II.
% general description of the disk mass evolution
\subsubsection{Disk Mass Evolution during the Class I Phase}
\begin{figure*}[htbp]
\centering
\includegraphics[width=1.8\columnwidth]{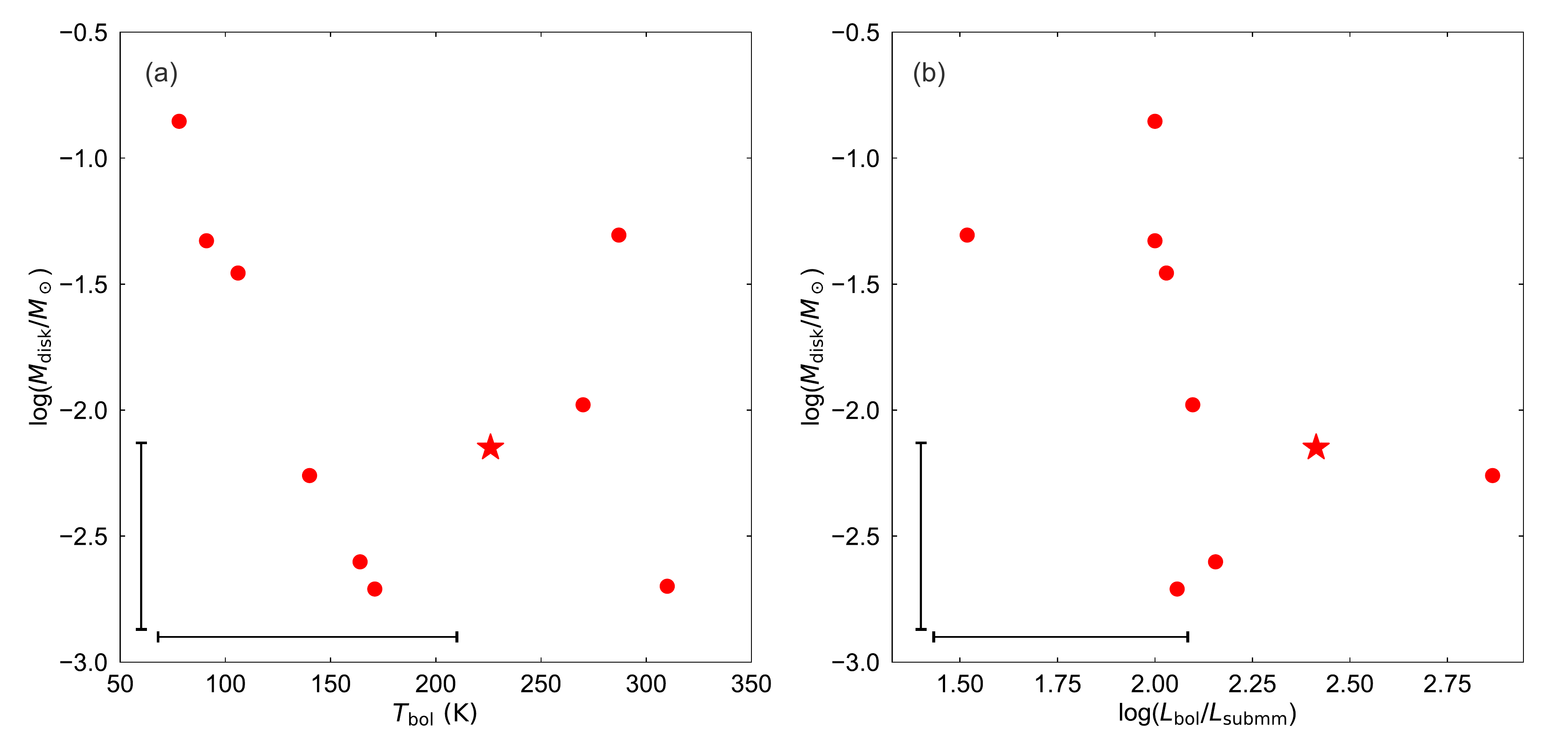}
\caption{Comparison between (a) $T_\mathrm{bol}$ and $M_\mathrm{disk}$ and (b) $L_\mathrm{bol}/L_\mathrm{submm}$ and $M_\mathrm{disk}$ among the Class I samples (listed in Table \ref{table_ClassIs}). Vartical bars at the bottom-left corners denote a typical uncertainty of $M_\mathrm{disk}$, which is derived from uncertainties of disk temperature, flux calibration error of $20\%$, and the index of the dust opacity $\beta$ through the error propagation. Horizontal bars at the bottom-left corners denote typical uncertainties of $T_\mathrm{bol}$ and $L_\mathrm{bol}/L_\mathrm{submm}$ \citep{Chen:1995aa}.}
\label{disk_corr}
\end{figure*}

Figure \ref{disk_corr} compares $\tbol$ and $L_\mathrm{bol}/L_\mathrm{submm}$ to $M_\mathrm{disk}$ among the Class I sources, the formers of which are widely-used indicators of evolutionary stages of young stellar objects (YSOs) \citep{Green:2013aa}. Typical uncertainties of the parameters are shown at the bottom-left corners in the figure. We assume the typical uncertainties of $\tbol$ and $L_\mathrm{bol}$ for embedded sources to be a factor of 1.5 and 2, respectively, as discussed in \cite{Chen:1995aa}. Note that these errors could be overestimated because these values are derived based on not only error of photometry but also the incompleteness of SED and the effect of extinction. We adopt the uncertainty of $L_\mathrm{bol}$ as that of $L_\mathrm{bol}/L_\mathrm{submm}$. The errors of $M_\mathrm{disk}$ are derived from the typical uncertainties of the disk temperature, the flux density of the dust continuum emission, and the dust mass opacity through the error propagation. We assume that the error of the disk temperature to be $\pm10$ K, and the error of the flux calibration of dust continuum emission to be $20\%$. Recent interferometric observations at submm and mm wavelengths suggest that the index $\beta$ of the dust mass opacity is typically less than unity even in Class 0 or Class I sources \citep{Jorgensen:2007aa,Kwon:2009aa,Miotello:2014ab}, whereas a couple of Class 0 sources shows $\beta$ close to the ISM value of $1.7$ \citep{Li:2017aa}. On the other hand, the averaged value of the index $\beta$ of Class \II disks are reported to be 0.6 in 21 sources in the Taurus-Auriga star-forming region, and 0.46 in 17 sources in the $\rho$ Ophiuchi young cluster \citep{Ricci:2010aa,Ricci:2010ab}. Thus, we assume the range of the $\beta$ in the Class I sample to be $1\pm0.5$.

Figure \ref{disk_corr}a shows a marginal, slight negative trend (the correlation coefficient is $-0.43$) even though the scattering and the errors of parameters are large. A similar mild negative trend could also be seen in Figure \ref{disk_corr}b (the correlation coefficient is $-0.49$). These possible trends may imply that $M_\mathrm{disk}$ decreases as disks evolve. \cite{Tychoniec:2018ab} reported a similar negative trend in the correlation between $\tbol$ and the disk masses of Class 0 and I sources with statistically sufficient samples. Our result could support the decline of $M_\mathrm{disk}$ with its evolution in an accretion phase. The disk mass range of our sample is 2.0$\times 10^{-3}$--0.14$\Msun$ after correcting for the opacity difference. This means that most of the disk mass $\sim$0.1$\Msun$ should be removed during the Class I phase. $M_\mathrm{disk}$ of the most evolved Class I sources are $\sim$(2--5)$\times 10^{-3} \Msun$, comparable to the average mass of Class \II disks in five different star forming regions, $\sim$1.5$\times 10^{-4}$--4.5$\times 10^{-3} \Msun$ on the assumption of the gas-to-dust mass ratio 100 \citep{Ansdell:2017aa}. Observational measurements of disk masses around Class 0 and I sources in the Perseus molecular cloud present the median $M_\mathrm{disk}$ values of 0.073 and 0.033$\Msun$, respectively \citep{Tychoniec:2018ab}.

% What reduces disk mass?
What removes most of $M_\mathrm{disk}$ during the Class I phase? 
% ########## Figure: Correlation between parameters ###
% #############################################
Early planet formation during the Class 0 or I stages has been suggested \citep[e.g.][]{Greaves:2010aa}. The mass that should be removed from Class I disks $\sim$0.03--0.1$\Msun$ is, however, larger than the minimum mass solar nebula (MMSN) and ten times the mass of Jupiter, implying that mechanisms other than planet formation are required to reduce $M_\mathrm{disk}$ in the Class I phase.

% mass accretion to the central star
% L1489 IRS
The simplest way to reduce $M_\mathrm{disk}$ is mass accretion to central stars. The accretion rate in each source is estimated from $\lbol$ and $M_\ast$ with the relation $\macc = \sqrt{R_\ast \lbol/G M_\ast}$, assuming that $\lbol$ is due to the accretion energy. The derived $\macc$ with $R_\ast=3 R_\odot$ is listed in Table \ref{table_ClassIs}. The typical value of $\macc$ is a few $\times 10^{-7} \msunpyr$ for Class I sources. On the other hand, disks around protostars gain mass through infall from envelopes. Infall rates $\minf$, estimated in some of the Class I sources, are typically a few $\times 10^{-6} \msunpyr$ and always larger than $\macc$, although the sample number is limited (Table \ref{table_ClassIs}). This fact means that $M_\mathrm{disk}$ is supposed to increase with time, contradicting the trend that $M_\mathrm{disk}$ decreases with evolution. Such a result, i.e., $\macc < \minf $ , has also been reported in Class 0 protostars and intermediate and high-mass (proto)stars, even though the sample size of Class 0 sources is small \citep{Beltran2016aa}. $\macc < \minf $ is also measured in HL Tau, which is at a transition from the Class I to Class \II phases \citep{Yen:2017ab}.

One possible explanation for the decrease of $M_\mathrm{disk}$ in total is episodic accretion. Disks continue to obtain mass from envelopes during the Class 0 and I phases and thus become gravitationally unstable. The mass that Class I disks are supposed to obtain within their lifetime is estimated to be $\sim$0.1--1$\Msun$, based on $\minf$ and the lifetime of Class I protostars $\sim 10^{5} \ \mathrm{yr}$ \citep{Wilking:1989aa,Kenyon:1990aa,Evans:2009aa,Dunham:2015aa}, suggesting that Class I disks can be gravitationally unstable and cause accretion bursts several times during their lives. Episodic accretion during the Class I phase is also suggested from protostellar evolutionary models \citep{Enoch:2009aa}. The disk mass is expected to decrease with evolution if $\minf$ declines with time because disks do not have time enough to be massive at later evolutionary stages. Infall rates likely decrease with time because infall should stop eventually at the Class \II stage. In fact, some observations report higher $\minf$ in several Class 0 sources $\sim 10^{-5} \msunpyr$ than in Class I sources \citep{Kristensen:2012aa,Mottram:2013aa}. Hence, episodic accretion with a decline of the infall rate possibly produces the observed trend that disk masses decrease during the Class I phase. It could also contribute to the relatively wide spread of $M_\mathrm{disk}$ in the plots.

% infall rate vs accretion rate
Another possible explanation for the trend in the disk mass is the change of the dust absorption opacity due to grain growth. The masses of evolved disks can be underestimated if the disks indeed have large grains that cannot be observed. The dust absorption opacity at mm wave-lengths increases with the dust size, while it decreases in the case of grains larger than mm-level \citep{Draine:2006aa,Ricci:2010aa}. This means that grain growth from mm size to around 10 cm is required to reduce the nominal Class I disk mass measured from mm emission by an order of magnitude. However, such large dust grains suffer radial drift and most of them fall onto central stars within $\sim10^5$ yr, although the efficiency of radial drift highly depends on the dust-to-gas mass ratio in disks \citep{Brauer:2008aa}. Moreover, envelopes would continue supplying small grains to disks. For these reasons, the change of dust absorption opacity due to the grain growth would not be feasible as the origin of the decrease of the disk mass.

On the other hand, recent works point out that the dust scattering becomes dominant in the dust opacity in protoplanetary disks with mm size dust grains, and it explains observational results at mm wave-lengths \citep{Zhu:2019aa,Carrasco-Gonzalez:2019aa}. Ignoring the dust scattering results in underestimating disk masses from the flux density at mm wave-lengths. Especially in compact disks with small radii ($r<30$ au), the disk masses can be underestimated by a factor of 10 \citep{Zhu:2019aa}. Hence, the dust scattering could also explain the decrease of the disk mass with evolution if the dust grains grow from $\micron$ size to mm size during the Class I stage. More sophisticated treatment of the dust opacity is needed to measure disk masses accurately.
%
% --------------- SECTION 6: Summary ----------------------
\section{Summary\label{summary}}
We have observed the Class I protostar L1489 IRS at a high spatial resolution of $\ang[angle-symbol-over-decimal]{;;0.3} \times \ang[angle-symbol-over-decimal]{;;0.2}$ with ALMA in the 1.3 mm continuum and the \ce{C^18O} $J=$2--1 line to investigate its spatial and kinematic structures on a scale of $\sim$40 au. The main results and conclusions can be summarized as follows:
% itemize
\begin{enumerate}
% continuum result
\item We detected that the 1.3 mm continuum emission originated from the circumstellar disk. The overall structure and brightness are consistent with those of previous ALMA observations, while the structure was more spatially resolved. We derived the total disk mass as $\sim$7.1$\times 10^{-3} \Msun$ from the measured total flux density. Assuming a geometrically thin disk, the inclination angle of the disk was estimated as $\sim$73\degree.
% Overall C18O results
\item The \ce{C^18O} $J=$2--1 line emission shows a clear velocity gradient from NE to SW, strong non-axisymmetry, and warping. The overall structure is consistent with that of the previous ALMA observations, but shows non-axisymmetry more strongly.
% Dips
\item The \ce{C^18O} emission shows dips at $r\sim280$ au on the NE side and $r\sim 140$ au on the SW side of the disk. The \ce{C^18O} dips are unlikely due to a real gas absence because the 1.3 mm continuum emission does not show any dips. Depletion of \ce{CO} molecules is a plausible source of the \ce{C^18O} dips. However, the physical cause of such a molecular depletion at certain disk radii is not clear.
% warped feature and rotation curve
\item The velocity gradient, which is seen from NE to SW in the \ce{C^18O} emission due to rotation, changes its position angle by $\sim$15\degree at radii where the \ce{C^18O} dips are found. This result suggests that the direction of the disk rotational axis differs between the inner and outer regions of the disk. To reveal the kinematic property of this warped structure, we examined the rotational profiles in detail. The best fit power-law index for the rotational profile inside the \ce{C^18O} dips is $0.50$, indicating Keplerian rotation. From the Keplerian rotation, the dynamical mass is estimated to be $\sim$1.6$\Msun$ with an inclination angle of $i=73\degree$, which is consistent with previous works. A double power-law function is, on the other hand, well fitted to the rotational profile outside the \ce{C^18O} dips, giving the best fit power-law indices of $p_\mathrm{in}=0.47$ and $p_\mathrm{out}=1.1$ with the break radius $R_\mathrm{break} \sim 600$ au. This result indicates that the warped structure $\lesssim$600 au is a Keplerian disk while the outer part is an infalling envelope.
% warped disk model and possible formation scenario
\item We constructed warped and coplanar disk models with radii of 600 au and compared them to the observations. The warped disk model, with its breaking radius of 200--300 au, shows a good agreement with the observations in both the channel maps and the PV diagram while the coplanar disk model exhibits offset in the P.A. direction from the observations in the channel maps. These results suggest that L1489 IRS is explained better as a warped disk system. On the other hand, there is extended emission outside 600 au, which seems to arise from the infalling envelope. The extended emission corresponds to arm-like emission interpreted as two infalling flows accreting onto the disk plane in the previous observations at a lower angular resolution. An accretion having an angular momentum axis in the different direction from that of the disk could be the origin of the warped structure. We performed a simple hydrodynamics simulation calculating disk formation through accretion where an angular momentum vector of the infalling gas varies its direction at a certain time during the simulation. The simulation demonstrates that such an accretion process can form a warped disk structure.
% Disk Mass and Radius of L1489 IRS
\item Comparing disk properties of Class I sources, including L1489 IRS, with those of Class \II sources, we found that gas disk radius distributions of Class I and \II disks are similar. On the other hand, the disk masses around Class I sources are higher than that around Class \II sources, even though the sample number is limited. L1489 IRS follows the trend of Class \II sources, which implies that the smaller disk mass of L1489 IRS is due to its evolutionary stage, i.e., a later stage of the Class I phase, regardless of its larger disk size than that of most other protostars.
% General Disk Evolution
\item Weak negative trends between $\tbol$ and $M_\mathrm{disk}$ and between $L_\mathrm{bol}/L_\mathrm{submm}$ and $M_\mathrm{disk}$ are seen among the Class I source sample, which could support the recent suggestion that the disk mass decreases even during the accretion phase. However, infall rates are always larger than accretion rates by an order of magnitude in YSOs, suggestive of increase of the disk mass as the disk evolves. Episodic accretion and decline of infall rates with evolution could reconcile an observed possible trend of the disk mass decreases as it evolves.
\end{enumerate}
%
% Acknowledgment
\acknowledgments
This paper used the following ALMA data: ADS/JAO.ALMA \#2013.1.01086.S (P.I. S. Koyamatsu). ALMA is a partnership of ESO (representing its member states), NSF (USA), and NINS (Japan), together with NRC (Canada), MOST and ASIAA (Taiwan), and KASI (Republic of Korea), in cooperation with the Republic of Chile. The Joint ALMA Observatory is operated by ESO, AUI/NRAO, and NAOJ. We thank all ALMA staff for conducting the observations. We also thank the anonymous referee, who gave us insightful comments and suggestions on improving the paper. Data analysis was in part carried out on the Multi-wavelength Data Analysis System operated by the Astronomy Data Center (ADC), National Astronomical Observatory of Japan. We thank M. Saito for the advice on data analysis. We thank M. Yamaguchi for the fruitful discussion. J.S. was supported by the Subaru Telescope Internship Program, and is supported by Academia Sinica Institute of Astronomy and Astrophysics. S.T. acknowledges a grant from JSPS KAKENHI Grant Number JP18K03703 in support of this work. Y.A. is grateful for NAOJ ALMA scientific Research Grant Number of 2019-13B. H.-W.Y. acknowledges support from MOST 108-2112-M-001-003-MY2. This work was supported by NAOJ ALMA Scientific Research grant No. 2017-04A.
\facility{ALMA}
%\facilities{facility ID, facility ID, facility ID} 
\software{\texttt{CASA} \citep{McMullin:2007aa}, \texttt{MIRIAD} \citep{Sault:1995aa}, \texttt{RADMC-3D} \citep{Dullemond:2012aa}, \texttt{Numpy} \citep{Oliphant:2006aa,van-der-Walt:2011aa}, \texttt{Scipy} \citep{Jones:2001aa}, \texttt{Astropy} \citep{Astropy-Collaboration:2013aa,Astropy-Collaboration:2018aa}, \texttt{Matplotlib} \citep{Hunter:2007aa}, \texttt{SFUMATO} \citep{Matsumoto:2007aa}}
%
% ################### APPENDIX ##########################
\appendix
% ########## Matsumoto-san's simulation ###############
\section{Numerical Simulations \label{app_sim}}The model presented here was constructed by extending that of \cite{Matsumoto:2019aa}. In the model here, a sink particle is considered at the center of the computational domain of $x,y,z \in [-12 R_0, 12 R_0]^3$, where $R_0 (=200~\mathrm{au})$ denotes the scale length of the inner disk.  The sink particle mimics a protostar, and its mass is assumed to be $M_\star = 1.6\  M_\sun$.  The computational domain is filled with a low-density gas of $10^{-3} \rho_0$ at the initial stage.  

The gas is injected at the spherical boundary surface with a radius of $12R_0$.  On the surface, the injected gas has a constant density $\rho_0$ and a radial velocity of $v_r = [2 G M_\star/r - (j_\mathrm{inf}/R)^2]^{1/2}$, assuming free fall from a distance of infinity, where $r$ is the spherical radius, $R$ is the cylindrical radius, and $j_\mathrm{inf}$ is the specific angular momentum.  We assume that the gas has a spatially constant angular velocity of $\Omega_\mathrm{inf}$. The injected gas mimics a rigidly rotating envelope and cloud core, which have often been considered in protostellar collapse simulations \citep[e.g.,][]{Matsumoto:2003aa,Machida:2010aa}. The gas is therefore injected with a range of specific angular momenta $0\le j_\mathrm{inf}\le j_\mathrm{max}$, where $j_\mathrm{max} = (12R_0)^2 \Omega_\mathrm{inf}$. We specify the angular momenta of the injected gas by a corresponding centrifugal radius $R_{c,\mathrm{max}} = j_\mathrm{max}^2/(GM_\star)$. 

In the early phase of $0 \le t \le 10 T_d$, the gas with $R_{c,\mathrm{max}} = 200~\mathrm{au}$ is injected at the spherical boundary surface, where $T_d$ ($=2\pi(R_0^3/GM_\star)^{1/2}$) denotes the rotation period of the gas at $R_0$.  This gas forms a circumstellar disk with a radius of 200~au around the sink particle. In the later phase of $t > 10 T_d$, the injected gas has a high angular momentum of $R_{c,\mathrm{max}} = 600~\mathrm{au}$, and the rotation axis of the envelope is inclined from the $z$-axis towards the $x$-axis by $15^\circ$.  This gas forms the outer disk, which is inclined by $15^\circ$ with respect to the inner disk. 

We use \texttt{SFUMATO} \citep{Matsumoto:2007aa} with so-called fixed mesh refinement. The six grid levels are considered, and the coarsest and finest resolutions are $\Delta x_\mathrm{max} = 9.38\times10^{-2} R_0$ and $\Delta x_\mathrm{min} = 2.93\times10^{-3}R_0$, respectively. The hydrodynamical scheme has third-order accuracy in space with MUSCL and second-order accuracy in time with the predictor-corrector method. The numerical flux is obtained via a Roe-type scheme \citep{Roe:1981aa}. The sink radius is set to $r_\mathrm{sink} = 4 \Delta x_\mathrm{min} = 1.17\times 10^{2} R_0$ \citep{Matsumoto:2015aa}. An isothermal gas with a constant sound speed of $c=0.1 (GM_\star/R_0)^{1/2}$ is assumed. We ignore the self-gravity of the gas and magnetic field.  
% ####################################################
% #################### pv fitting evaluation ################
\section{Evaluation of Fitting Method on PV Diagram \label{app_pv}}
% figure
% Rn vs vpeak,n 
\begin{figure}
\centering
\includegraphics[viewport=100 0 742 595, width=0.5\columnwidth]{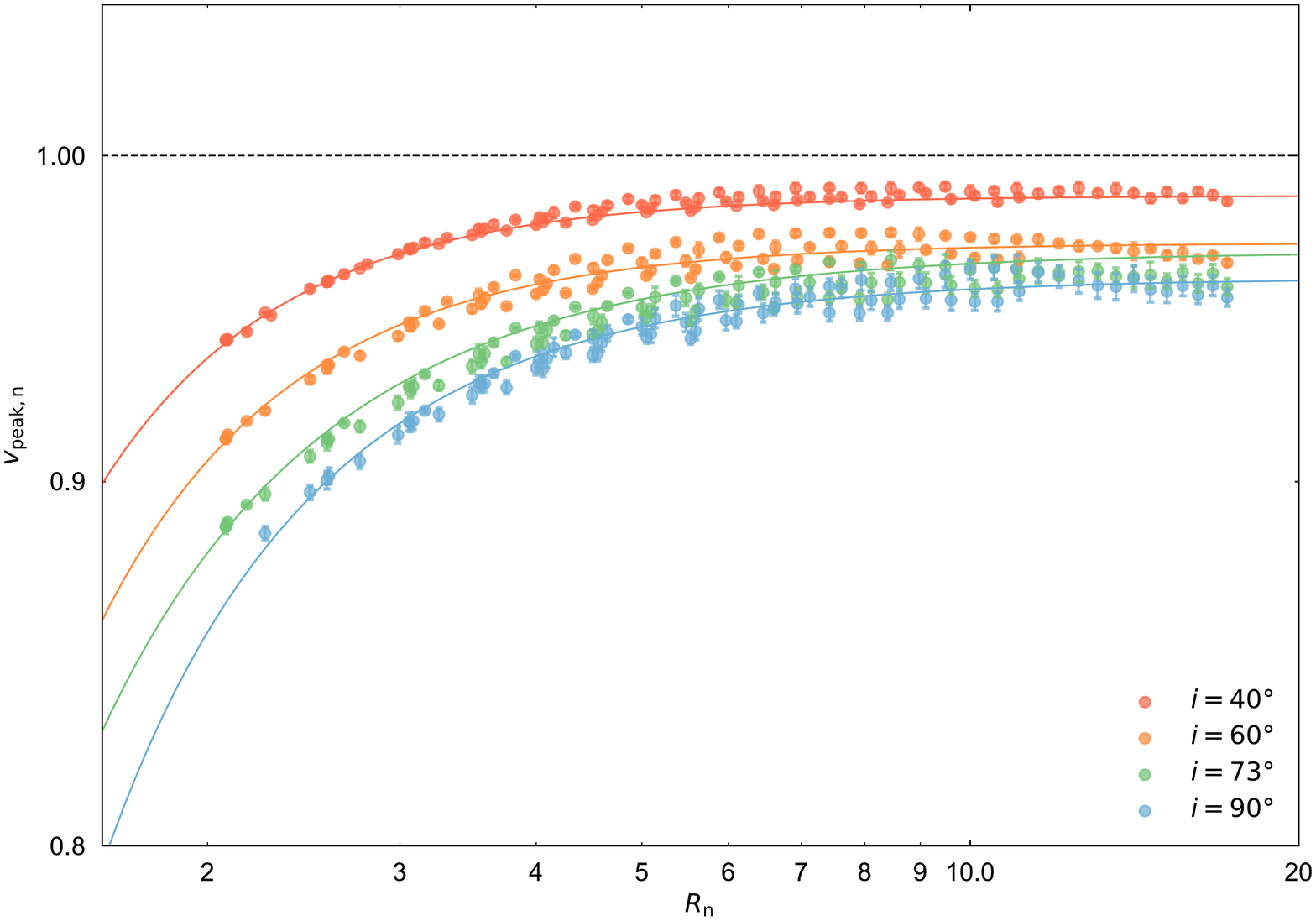} % 0 0 842 595
\caption{Normalized radius $R_\mathrm{n}=R/\theta$ vs normalized velocity $v_\mathrm{peak, n}=v_\mathrm{peak}/v_\mathrm{exp}$ determined by fitting on PV diagrams. All fitting results for the models with different angular resolutions are shown in the same color for each inclination angle. Solid lines show the best fit functions described by Equation (\ref{app_eq_fx}).}
\label{rn_vs_vn}
\end{figure}
%
% Rn vs vpeak,n to compare different dv
\begin{figure}
\centering
\includegraphics[viewport=100 0 742 595, width=0.5\columnwidth]{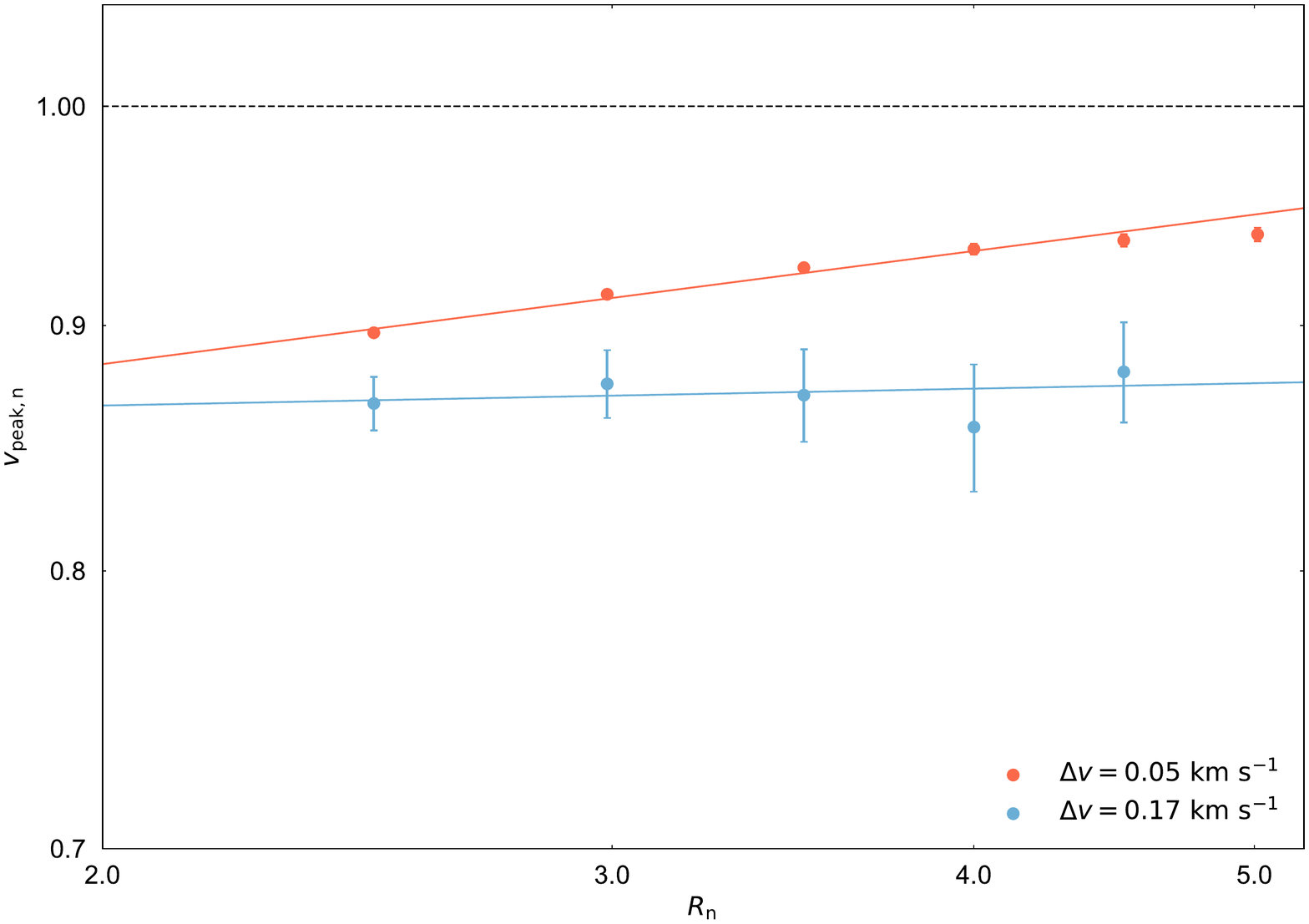} % 0 0 842 595
\caption{Normalized radius $R_\mathrm{n}=R/\theta$ vs normalized velocity $v_\mathrm{peak, n}=v_\mathrm{peak}/v_\mathrm{exp}$ for the case models with the same inclination angle $i=90\degree$ and beam size $\theta = \ang[angle-symbol-over-decimal]{;;0.7}$ but with different velocity resolutions of $0.05\ \kmps$ and $0.17\ \kmps$. Solid lines show the fitting results with a simple power-law function ($v_\mathrm{peak,n} \propto R_\mathrm{n}^{p}$).}
\label{rn_vs_vn_compdv}
\end{figure}
%%%%%%%%%%%%%%%%%%%%%%%%%%%%%%%%%%%%%%%%%%%%%%%%%%%%%%%%%%
In Section \ref{analysis_vc18o}, a new method to determine the representative velocities at large radii, which uses only five channels around the peak of a spectrum for Gaussian fitting, was introduced. Our motivation to introduce this new method was to minimize the effect of spectral skewing due to beam smearing and overlaps of multiple velocity components along the line-of-sight. These effects are significant, especially when the beam size and inclination angle of a source are large, resulting in an incorrect rotation profile. Here, we examine how correctly the new method derives the rotational profile.

We constructed Keplerian disk models, which are the same as the coplanar disk model in Section \ref{kineticmodel} but with different inclination angles of $40\degree, 60\degree, 73\degree$, and $90\degree$ and angular resolutions of $\ang[angle-symbol-over-decimal]{;;0.2}$ to $\ang[angle-symbol-over-decimal]{;;0.8}$ with steps of $\ang[angle-symbol-over-decimal]{;;0.1}$. The stellar mass and disk outer radius are assumed to be $1.6 \Msun$ and $600$ au, respectively. Velocity resolutions of the models are set at $0.05 \kmps$ to consider an ideal case where the spectral shapes are sufficiently resolved. The effect of the velocity resolution is discussed later. The same fitting method as in Section \ref{analysis_vc18o} is applied to the PV diagrams of the models, and the peak velocity $v_\mathrm{peak}$ is determined at radius $R$. We examine the dependence of the determined peak velocity normalized as $v_\mathrm{peak, n}=v_\mathrm{peak}/v_\mathrm{exp}$ on the normalized radius $R_\mathrm{n} = R/\theta$, where $v_\mathrm{exp}$ is the expected rotational velocity at the radius, and $\theta$ is the beam size. To cover a wide range of $R_\mathrm{n}$, the fitting is performed with the models with different angular resolutions within the range of $\ang[angle-symbol-over-decimal]{;;0.1} < R < \ang[angle-symbol-over-decimal]{;;3.4}$ for each inclination angle.
%%%%%%%%%%%%%%%%%%%%%%%%%%%%%%%%%%%%%%%%%%%%%%%%%%%%%%%%%%
% Table
%
\begin{table*}[htbp]
\centering
\begin{threeparttable}
    \begingroup
    \centering
    \caption{Best fit parameters for Equation (\ref{app_eq_fx}) and parameters derived from Equation (\ref{app_eq_Rnlim}).}
    \begin{tabular*}{0.8\columnwidth}{@{\extracolsep{\fill}} lccccc}
        \hline\hline
         $i$ & $v_0$ & $q$ & $c_0$ & $R_\mathrm{n, 10\%}$ \tnote{a} & $R_\mathrm{n, 20\%}$ \tnote{a} \\ \hline
         $40\degree$ & $0.282 \pm 0.014$ & $2.484 \pm 0.062$ & $0.98718 \pm 0.00027$ & 2.93 & 2.24 \\ 
         $60\degree$ & $0.338 \pm 0.018$ & $2.357 \pm 0.070$ & $0.97219 \pm 0.00048$ & 3.30 & 2.48 \\ 
         $73\degree$ & $0.351 \pm 0.016$ & $1.964 \pm 0.072$ & $0.9695 \pm 0.0012$ & 3.91 & 2.78 \\ 
         $90\degree$ & $ 0.449 \pm 0.076$ & $2.11 \pm 0.18$ & $0.9612 \pm 0.0021$ & 4.15 & 3.02 \\ \hline
    \end{tabular*}
    \begin{tablenotes}
	\item[a] $R_\mathrm{n, 10\%}$ and $R_\mathrm{n, 20\%}$ are the minimum radii of data points used for the fitting to achieve 10\% and 20\% accuracies for Keplerian rotation in the measurement of the rotational power-law index, which are derived from Equation (\ref{app_eq_Rnlim}) with $p_\mathrm{err} = 0.05$ and $p_\mathrm{err} = 0.1$, respectively.
	\end{tablenotes}
	\endgroup
    \end{threeparttable}
    \label{app_tab_fparams}
\end{table*}
%%%%%%%%%%%%%%%%%%%%%%%%%%%%%%%%%%%%%%%%%%%%%%%%%%%%%%%%

Figure \ref{rn_vs_vn} shows the relation between $R_\mathrm{n}$ and $v_\mathrm{peak, n}$. $v_\mathrm{peak, n}$ is less than unity at any radius, which means that the peak velocity of the spectrum does not correspond to the rotational velocity exactly. $v_\mathrm{peak, n}$ increases with radius gradually because the effects of beam smearing and integration along the line-of-sight are more significant at small radii. $v_\mathrm{peak, n}$ approaches a value less than unity at a large radius. This is because the effect of integration along the line-of-sight skews the line profile even if the disk is spatially resolved. Figure \ref{rn_vs_vn} also shows that the fitting to a disk with a high inclination angle results in a low $v_\mathrm{peak, n}$ value, which is also due to the integration effect. The radial dependence of $v_\mathrm{peak, n}$ indicates that the rotational profile derived from the determined peak velocity, which traces the spectral peaks better than the fitting without any channel selection, can be different from the genuine rotation profile of the disk even with this method.

To evaluate how the radial dependence of $v_\mathrm{peak, n}$ affects the derived rotation profile quantitatively, we fit the following function to $v_\mathrm{peak, n}$ and derive the slope of the function in the log-log plane from the derivative of the logarithm of the function:
\begin{align}
    v_\mathrm{peak, n} = - v_0 R_\mathrm{n}^{-q} + c_0,
    \label{app_eq_fx}
\end{align}
where $v_0$ and $q$ are positive values. The best fit results are shown in Figure \ref{rn_vs_vn}. $p_\mathrm{err} (R_\mathrm{n})$, which is the local slope of the function in the log-log plane, is derived from Equation (\ref{app_eq_fx}) as
\begin{align}
    p_\mathrm{err} & = \frac{d \log v_\mathrm{peak, n}}{d \log R_\mathrm{n}} \nonumber \\
    & = \frac{R_\mathrm{n}}{v_\mathrm{peak, n}} \frac{d v_\mathrm{peak, n}}{d R_\mathrm{n}} \nonumber \\
    & = \frac{v_0 q}{-v_0 + c_0 R_\mathrm{n}^{q} }.
    \label{app_eq_perr}
\end{align}
This $p_\mathrm{err}$ becomes an uncertainty of the derived rotational power-law index at $R_\mathrm{n}$ as $v \propto r^{-p_\mathrm{true} + p_\mathrm{err}}$, where $v_\mathrm{exp} \propto r^{-p_\mathrm{true}}$ and $p_\mathrm{true} = 0.5$ represent Keplerian rotation because $v_\mathrm{peak, n}$ locally follows $v_\mathrm{peak, n} \propto r^{p_\mathrm{err}}$ and now $v_\mathrm{peak} = v_\mathrm{peak, n} \times v_\mathrm{exp}$. From Equation (\ref{app_eq_perr}), we obtain the condition that $R_\mathrm{n}$ should satisfy to derive the rotational power-law index within an error of $p_\mathrm{err}$ as
\begin{align}
    R_\mathrm{n} > \left\{ \frac{c_0 p_\mathrm{err}}{v_0 (p_\mathrm{err} + q)} \right\}^{-1/q}.
    \label{app_eq_Rnlim}
\end{align}
For the case of the Keplerian rotation $v_\mathrm{exp} \propto r^{-0.5}$, the fitting to data points satisfying Equation (\ref{app_eq_Rnlim}) with $p_\mathrm{err} = 0.05$ derives the rotational profile with 10\% accuracy for Keplerian rotation. Note that $p_\mathrm{err}$ is an upper limit of the uncertainty because data points at larger radii than that exactly corresponding to $p_\mathrm{err}$ in Equation (\ref{app_eq_perr}) are included in the fitting. As seen in Figure \ref{rn_vs_vn}, a case with a higher inclination angle requires a larger limit of $R_\mathrm{n}$ than with a lower inclination angle to achieve the same $p_\mathrm{err}$.

In the case of L1489 IRS, the inclination angle of the outer disk is unclear. Hence, we assume $i=90\degree$ conservatively to evaluate $p_\mathrm{err}$. The best fit values are $(v_0, q, c_0) = (0.449 \pm 0.076, 2.11 \pm 0.18, 0.9612 \pm 0.0021)$ for the case with $i=90\degree$. The parameter errors are the fitting errors and shown as $3\sigma$. The error of $p_\mathrm{err}$ calculated from the parameter errors through the error propagation is less than 0.5\% in the range of $R_\mathrm{n}=$2--20 and is thus ignored here. With these values, 10\% and 20\% accuracies for Keplerian rotation (i.e., $p_\mathrm{err} = 0.05$ and $p_\mathrm{err} = 0.1$) require $R_\mathrm{n} > 4.15$, and $R_\mathrm{n} > 3.02$, respectively. In our analysis in Section \ref{analysis_vc18o}, the inner-most data point is $R_\mathrm{n} \sim 3.14$ with $R \sim \ang[angle-symbol-over-decimal]{;;2.2}$ and $\theta \sim \ang[angle-symbol-over-decimal]{;;0.7}$, suggesting the uncertainty is less than 20\% of the power-law index of Keplerian rotation. The results of the cases with the different inclination angles are listed in Table \ref{app_tab_fparams}.

Fitting on PV diagrams is also performed with the same models but with the velocity resolution of the current observations, $\Delta v = 0.17\ \kmps$, to investigate the effect of the velocity resolution. Figure \ref{rn_vs_vn_compdv} shows the correlation between $R_\mathrm{n}$ and $v_\mathrm{peak, n}$ for the models with the same inclination angle $i=90\degree$ and beam size $\theta = \ang[angle-symbol-over-decimal]{;;0.7}$ but with different velocity resolutions of $0.05\ \kmps$ and $0.17\ \kmps$. $v_\mathrm{peak, n}$ for the model with $\Delta v = 0.17\ \kmps$ is smaller than that for the model with $\Delta v = 0.05\ \kmps$ at any radius. The radial dependency of $v_\mathrm{peak, n}$ for the model with $\Delta v = 0.17\ \kmps$ is weaker than that for the model with $\Delta v = 0.05\ \kmps$. These results are because the spectrum shape is not sufficiently resolved in the model with a coarse velocity resolution and close to Gaussian shape even at small radii where the spectrum is skewed severely. To evaluate the difference of the radial dependencies quantitatively, a simple power-law function $v_\mathrm{peak, n} \propto R_\mathrm{n}^p$ is fitted to these data. The derived power-law indices are $p=0.01\pm0.053$ for the model with $\Delta v = 0.17\ \kmps$ and $p=0.08\pm0.015$ for the model with $\Delta v = 0.05\ \kmps$, indicating that $v_\mathrm{peak, n}$ for the model with $\Delta v = 0.17\ \kmps$ is close to constant at these radii, and thus the velocity resolution of the current observations does not increase the uncertainty of the derived power-law index expected in the ideal case. Therefore, the uncertainty of the derived rotational power-law index in addition to the fitting error in Section \ref{analysis_vc18o} is less than $0.1$ (20\% of the power-law index of Keplerian rotation). Note again that this value would be an upper limit because the fitting is performed using data points at larger radii than the inner-most point.
% ############## maps of the 13CO ########################
\section{\ce{^13CO} Line Emission \label{app_13co}}
\begin{figure}
\centering
\includegraphics[viewport=120 0 722 595, width=0.5\columnwidth]{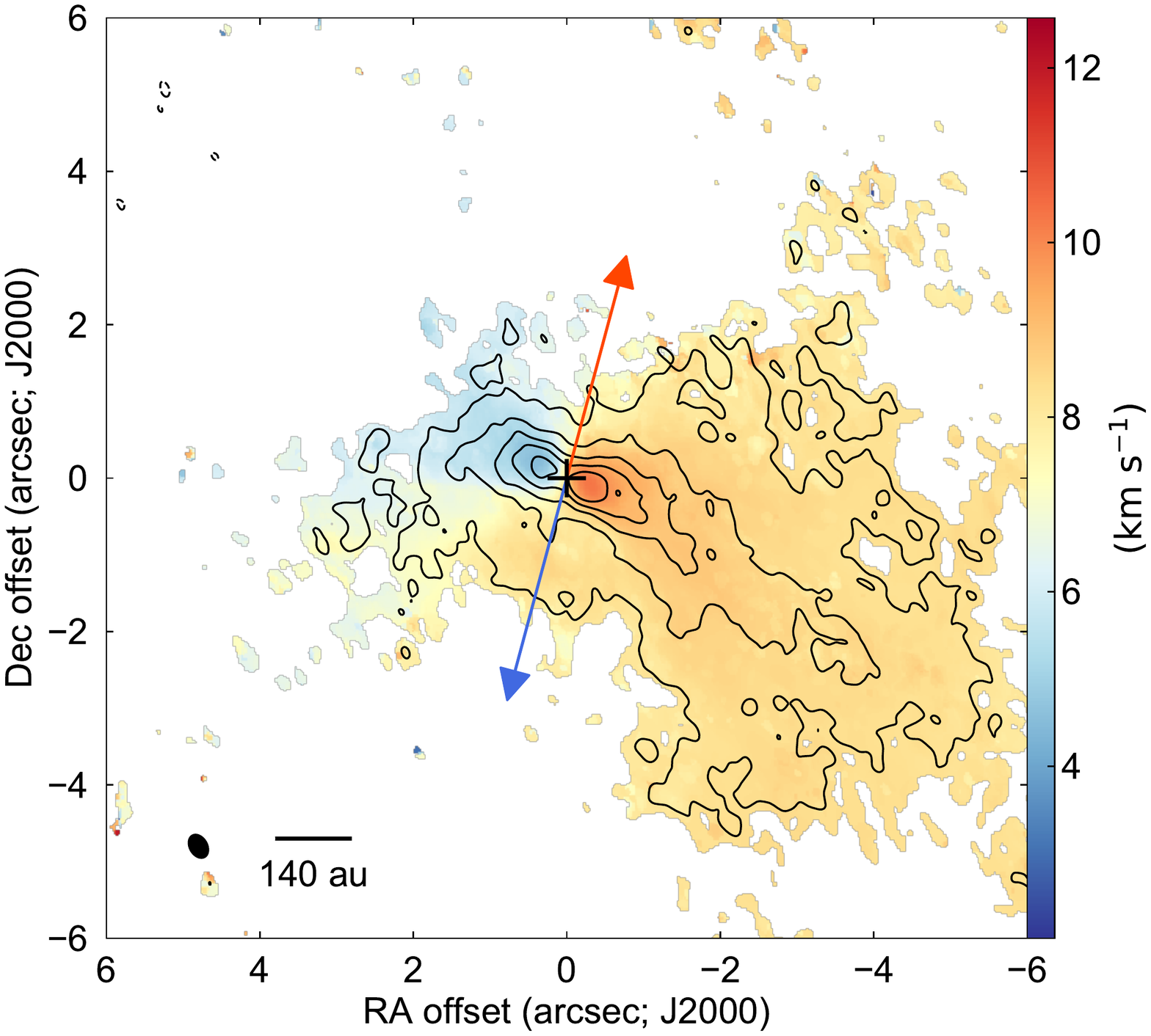} % 0 0 842 595
\caption{Total intensity map (\textit{contours}) and mean velocity map (\textit{color}) of the \ce{^13CO} $J=$2--1 emission of L1489 IRS. Contour levels are -3, 3, 6, 9, 12, 15 $\times \sigma$, where 1$\sigma$ corresponds to $16 \mjpbm$. Negative contours are shown with dashed lines. The cross shows the position of the central protostar. The blue and red arrows denote the direction of the blueshifted and redshifted components of the outflow, respectively \citep{Hogerheijde:1998aa}. The black-filled ellipse in the bottom-left corner indicates the synthesized beam size: $\ang[angle-symbol-over-decimal]{;;0.35}\times \ang[angle-symbol-over-decimal]{;;0.25}$, P.A.$=\ang{30}$.}
\label{mom13co21}
\end{figure}

\begin{figure*}
\centering
\includegraphics[width=\columnwidth]{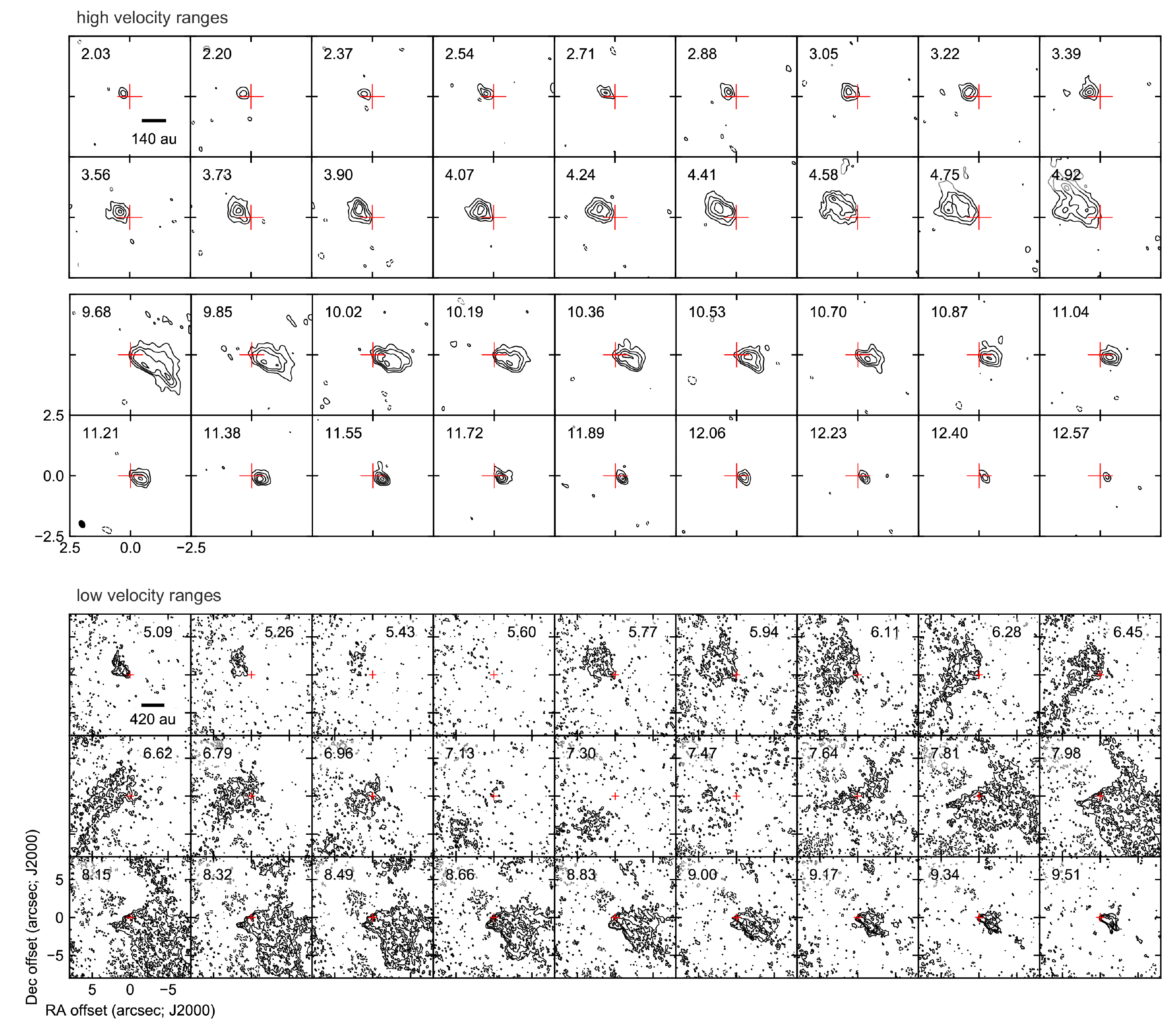}
\caption{Channel maps of the \ce{^13CO} $J=$2--1 emission at the higher (\textit{upper panels}) and lower velocity ranges (\textit{lower panels}). Note that the spatial coverage is distinct between the two panels in order to show the compact high-velocity emission more clearly. Contour levels are -3, 3, 6, 9, 12, 15 $\times \sigma$, where 1$\sigma$ corresponds to $5.9 \mjpbm$. Negative contours are shown with dashed lines. The LSRK velocity of each panel is indicated in the top-left (or top-right) corners in $\kmps$. The red cross at the center of each panel shows the position of the central protostar. The black-filled ellipses in the bottom-left corners indicate the synthesized beam size: $\ang[angle-symbol-over-decimal]{;;0.35}\times \ang[angle-symbol-over-decimal]{;;0.25}$, P.A.$=\ang{30}$.}
\label{channel13co21}
\end{figure*}

The total intensity map integrated over a velocity range of 2.03--12.57$\kmps$ overlaid on the mean velocity map and channel maps of the \ce{^13CO} $J=$2--1 line emission are shown in Figure\ref{mom13co21} and \ref{channel13co21}, respectively.
% ########## Figures: 13CO ############
% moment maps
%
% channel map
%
% ###### REFERENCE ######
\bibliographystyle{yahapj}
\footnotesize{\bibliography{reference}}
% #######################
\end{document}